\documentclass[useAMS,usenatbib]{mnras}     
\usepackage{times}
\usepackage{./aas_macros}
\usepackage{graphicx}
\graphicspath{{images/}}
\usepackage{amssymb}
\usepackage{amsmath}
\usepackage{xcolor}
\usepackage{multirow}
\usepackage{soul}

\newcommand{\D}[0]{{\mathcal D}}
\newcommand{\M}[0]{{\mathcal M}}
\newcommand{\V}[0]{{\mathcal V}}

\begin{document}

%
%\title{On the route to constrain properties of high density matter in 
%neutron stars with magneto-elastic oscillations}
\title{Constraining properties of high-density matter in neutron stars with magneto-elastic oscillations}
\author[Michael Gabler, Pablo Cerd\'a-Dur\'an, Nikolaos Stergioulas, Jos\'e A.~Font and Ewald M\"uller]
{Michael Gabler$^{1}$, %\thanks{E-mail:miga@mpa-garching.mpg.de}, 
Pablo Cerd\'a-Dur\'an$^2$, %\thanks{E-mail:cerda@mpa-garching.mpg.de},
Nikolaos Stergioulas$^3$, %\thanks{E-mail:niksterg@auth.gr}
Jos\'e A.~Font$^{2,4}$, %\thanks{E-mail:j.antonio.font@uv.es}
\and and 
Ewald M\"uller$^1$ %\thanks{E-mail:emueller@mpa-garching.mpg.de}
\\
  $^1$Max-Planck-Institut f\"ur Astrophysik,
  Karl-Schwarzschild-Str.~1, 85741 Garching, Germany \\
  $^2$Departamento de Astronom\'{\i}a y Astrof\'{\i}sica,
  Universidad de Valencia, 46100 Burjassot (Valencia), Spain\\
  $^3$Department of Physics, Aristotle University of Thessaloniki,
  Thessaloniki 54124, Greece \\
  $^4$ Observatori Astron\`omic, Universitat de Val\`encia, C/ Catedr\'atico 
  Jos\'e Beltr\'an 2, 46980, Paterna (Val\`encia), Spain \\
  }
\date{\today}
\maketitle
\begin{abstract}
We discuss torsional oscillations of highly magnetised neutron stars (magnetars) 
using two-dimensional, magneto-elastic-hydrodynamical simulations. Our model is 
able to explain both the low- and high-frequency quasi-periodic oscillations 
(QPOs) observed in magnetars. The analysis of these oscillations provides 
constraints on the breakout magnetic-field strength, on the fundamental QPO 
frequency, and on the frequency of a particularly excited overtone. More 
importantly, we show how to use this information to generically constraint  
properties of high-density matter in neutron stars, employing Bayesian analysis.
In spite of current uncertainties and computational 
approximations, our model-dependent Bayesian posterior estimates for SGR 
1806-20 yield a magnetic-field strength $\bar B\sim 
2.1^{+1.3}_{-1.0}\times10^{15}\,$G and a crust thickness of $\Delta r = 
1.6^{+0.7}_{-0.6}$ km,
which are both in remarkable agreement with 
observational and theoretical expectations, respectively (1-$\sigma$ 
error bars are indicated). Our posteriors also favour the presence of a 
superfluid phase in the core, a relatively low stellar compactness, 
$M/R<0.19$, indicating a relatively stiff equation of state and/or low 
mass neutron star, and high shear speeds at the base of the crust, 
$c_s>1.4\times10^8\,$cm/s. Although the procedure laid out here still 
has large uncertainties, these constraints could become tighter when additional 
observations become available.
\end{abstract}
%

%=============================================================================
\section{Introduction}
The quasi-periodic oscillations (QPOs) observed in the X-ray tail of  so-called giant flares in magnetars \citep[e.g.][]{Israel2005, Strohmayer2005, Watts2006,Strohmayer2006} have stimulated theoretical research in neutron star oscillations. 
QPOs have been interpreted  as toroidal shear oscillations of the solid crust by various groups, and approaches to constrain the nuclear equation of state (EoS) of the crust have started \citep[see][and references therein]{Duncan1998, 
Messios2001, Strohmayer2005, Piro2005, Sotani2007, Samuelsson2007, Steiner2009, Samuelsson2009, Sotani2013, Deibel2014, Sotani2016}. However, since the neutron stars where QPOs are observed have the strongest 
magnetic fields known, it was realized that the interaction of the crust with the magnetic field leads to a very effective coupling of the two \citep{Levin2006,Levin2007,Glampedakis2006b}. Detailed theoretical models of both, 
Alfv\'en oscillations \citep{Cerda2009, Sotani2008, Colaiuda2009} and coupled magneto-elastic oscillations\citep{Gabler2011letter, Gabler2012, Colaiuda2011, vanHoven2011, vanHoven2012} have been extended to even include the effects of superfluid neutrons which are expected in the neutron star core \citep{vanHoven2011, vanHoven2012, Glampedakis2011a, Passamonti2013, Gabler2013b, Passamonti2014, Gabler2016}.

However, the magneto-elastic model has problems to identify all of the observed QPO frequencies. These can be divided into two groups, each family with frequencies either below or above $260\,$Hz. The low-frequency QPOs confirmed by different 
groups are: $18$, $26$, $30$, $92$, $150$ (giant flare of SGR 1806-20), and $28$, $53$, $84$, $155$\,Hz (giant flare of SGR 1900+14) \citep[see e.g.][]{Israel2005, Strohmayer2005, Watts2006,Strohmayer2006}. Other studies found 
additional possible oscillations at $17$, $21$, $36$, $59$, and $116\,$Hz \citep[giant flare of SGR 1806-20,][]{Hambaryan2011}, $57\,$Hz  \citep[less energetic bursts of SGR 1806-20,][]{Huppenkothen2014b},
and $93$, $127$, and $260\,$Hz \citep[less energetic bursts of SGR J1550-5418,][]{Huppenkothen2014a}. The corresponding high-frequency QPOs are $625\,$Hz and  $1840\,$Hz, and have been found in the giant flare of SGR 1806-20. The first magneto-elastic models that succeeded in explaining both low- and high-frequency QPOs were presented by \cite{Gabler2013b} and confirmed by \cite{Passamonti2014}.

The identification of the oscillation mechanisms responsible for the observed 
QPO frequencies would allow to constrain the properties of the high-density 
matter in the hardly accessible interior of neutron stars. In addition to 
improving our knowledge of superfluid parameters, crustal composition, and core 
composition, one could also gain information about the strength of the interior 
magnetic field as well as about its configuration. Enormous progress has been 
achieved in characterising  pure crustal shear modes depending on the equation 
of state (EoS) \citep{Steiner2009, Sotani2011, Sotani2012, Sotani2013, 
Sotani2013b, Sotani2016}. However, these models completely neglect the presence 
of the magnetic field and the resulting coupling between the crust and the core. 
For the more complicated problem of global magneto-elastic oscillations, only 
very general constraints on the neutron star EoS and/or magnetic-field 
properties have thus far been obtained \citep{Sotani2008, Andersson2009, 
Colaiuda2011, Gabler2013b}. In this work we take a step forward in this 
direction and show how a Bayesian analysis of magneto-elastic oscillations of 
magnetars may help to constrain different properties of high-density matter in 
neutron stars.

The organisation of this paper is as follows: in Section\,\ref{sec_theory} we present a summary of our theoretical model based on the work by \cite{Gabler2016}. In Section~\ref{sec_model} we describe three observational predictions of our theoretical model that can be used to constrain properties of neutron star matter.  The corresponding constraints are discussed in Section~\ref{sec_constraints}. Finally, our conclusions are presented in Section\,\ref{sec_conclusion}.

%
%=============================================================================

\section{Theoretical framework}\label{sec_theory}

We follow \cite{Gabler2016} and solve the general-relativistic magneto-elastic-hydrodynamical equations with our numerical code {\tt MCOCOA}  \citep[see also][for details]{Cerda2008, Cerda2009, Gabler2011letter, Gabler2012, Gabler2013a}. 
We use a spherically symmetric metric in isotropic coordinates that can be derived from the line element 
\begin{equation}
 ds^2 = - \alpha^2 dt^2 + \Phi^4\hat\gamma_{ij} dx^i dx^j \,,
\end{equation}
where $\alpha$, $\Phi$, and $\hat\gamma_{ij}\equiv{\rm diag}\left(1,r^2,r^2 \sin\theta\right)$ are the lapse function, the conformal factor, and the spatial flat 3-metric, respectively. The circumferential radius $R$ is thus related to the coordinate radius $r$ by $R=\Phi^2 r$. For our simulations we neglect the dynamics of the spacetime (Cowling approximation).

The equations describing torsional magneto-elastic oscillations are a consequence of the conservation of baryon number, energy, momentum, and Maxwell's equations. We only consider linear perturbations in axisymmetry. In this case, poloidal and toroidal perturbations decouple and the system of equations can be cast as \citep{Gabler2011letter, Gabler2012, Gabler2016}:
\begin{equation}
 \frac{1}{\sqrt{-g}} \left( \frac{\partial\sqrt{\gamma} \mathbf{U}
}{\partial t} +
\frac{\partial \sqrt{-g} \mathbf{F}^i}{\partial x^i} \right) = 0\,,
\label{conservationlaw}
\end{equation}
where $x^i\equiv (r,\theta)$, $g$ is the determinant of the 4-metric, $\gamma$ is the determinant of the 3-metric, and the state vector $\bf U$ and the flux vectors $\bf F^i$ are given by:
\begin{eqnarray}
 \mathbf{U} &=& [S^{(c)}_\varphi, B^\varphi]  \label{reduced_withcrust1_c}\,,\\
 \mathbf{F}^r &=& \left[ -
\frac{b_\varphi B^r}{W^{(c)}} - 2 \mu_\mathrm{S}
\Sigma^r_{~\varphi}, - v^{\varphi(c)} B^r
\right]\,,  \label{flux_r_p}\\
 \mathbf{F}^\theta &=& \left[ - \frac{b_\varphi B^\theta}{W^{(c)}}- 2
\mu_\mathrm{S} \Sigma^\theta_{~\varphi},
-v^{\varphi(c)} B^\theta
\right]\,.\label{flux_theta_p}\label{reduced_withcrust2_c}
\end{eqnarray}
Here $B^i$ is the magnetic field measured by an Eulerian observer while $b_i$ is that measured by a co-moving observer, $\Sigma^{\mu\nu} = g^{\mu\mu} \xi^{\nu(c)}_{,\mu}$ is the shear tensor, and $\mathbf{\xi}^{(c)}$ is the displacement ($\xi^{j(c)}_{\,,t} = \alpha v^{j(c)}$). The Lorentz factor $W^{(c)}=\alpha u^{t(c)}$, the three-velocity $v^{i(c)}$, and the generalized momentum density $S^{(c)}_i=(\varepsilon_\star X_c \rho h + b^2) W^{(c)2} v^{(c)}_i - \alpha b_i b^0$ are those of 
the {\it charged} particles only (protons in this work). Note that superfluid effects are introduced by the mass fraction of charged particles $X_c$ and by the entrainment factor $\varepsilon_\star  = (1-\varepsilon_n)/(1-\varepsilon_n-\varepsilon_c)$, where $\varepsilon_n$ and $\varepsilon_c$ are the entrainment coefficients of neutrons and the charged component, respectively. In this work we use the combination of $X_c \varepsilon_\star$ as a parameter to study possible effects of superfluidity on the oscillations.

To solve the evolution equations we have to provide boundary conditions at the 
surface of the star. There, the continuous traction condition has to be 
fulfilled which leads to 
\begin{eqnarray}
 b^\varphi_\mathrm{crust} &=& b^\varphi_\mathrm{atmosphere}\,,\\
\xi^{\varphi(c)}_\mathrm{crust,r}&=&0\,.
\end{eqnarray}
At the core-crust interface our MHD solver ensures the continuity of the 
momentum and can cope with the appearing discontinuities. Therefore, we do not 
apply any additional conditions there. 
We initiate the simulations with a general perturbation consisting of the 
overlap of several spherical harmonics, which should excite many of the 
magneto-elasticoscillations. 

Within this effective one-fluid approach we neglect additional effects that arise due to presence of superconducting protons which change the MHD equations, superfluid vortices which may couple to the crust by pinning or which may couple the superfluid neutrons to the charged components by scattering protons off quasi-particles confined in the cores of neutron 
vortices via the strong (nuclear) force \citep{Sedrakian2016}. We also neglect any effects arising due to the slow rotation of magnetars. For a detailed discussion we refer the interested reader to \cite{Gabler2016}

The magneto-elastic framework we have just described allows to perform numerical simulations of the evolution of torsional 
oscillations of perturbed neutron stars. However, the associated space of parameters of the framework is fairly large and its investigation is computationally expensive. The set of parameters include the EoS, the mass of the neutron star, the magnetic field strength and structure, the entrainment parameter ($\varepsilon_\star$), the proton fraction ($X_c$) and the shear modulus ($\mu_{\rm S}$). Numerical simulations enable to extract the QPO frequencies corresponding to any particular set of parameters. During the last few years the parameter space has been extensively explored and relations between the QPO frequencies and the different parameters have been extracted. In the current work we use the results  provided by \cite{Sotani2008,Cerda2009,Gabler2011letter,Gabler2012, Gabler2013a, Gabler2013b, Gabler2016}. We parameterise the strength of the magnetic field by $\bar B$, the {\it equivalent dipole magnetic field} \citep[see][]{Gabler2013a}, that corresponds to the magnetic field of a uniformly magnetised sphere rotating rigidly with a radius of $10\,$km and the same dipolar magnetic momentum as our model. This definition allows to compare models with different magnetic-field structure that may have the same dipolar momentum (so their spin-down properties are similar) but a very different surface magnetic-field strength. Moreover, to simplify the analysis, we parameterise the combination $X_c\varepsilon_\star$ (instead of the individual parameters) to study the effects of superfluidity on the oscillations. This has the advantage to allow us to simulate different conditions without knowing the detailed physical processes that determine the interaction of protons, electrons and neutrons that may or may not be superfluid or superconducting, respectively. Furthermore, using this approach, the different entrainment and interaction processes, like the mutual entrainment, interaction between neutron vortices with normal protons or proton flux tubes, and with the crustal lattice due to pinning, can be estimated. 

To complete the exploration of the parameter space performed in previous 
studies, we also discuss in this work additional simulations. Each of the new 
models (series N hereafter) has been constructed as a stratified fluid 
equilibrium configuration using a modified 
version of the RNS code \citep{Stergioulas1995}, which was extended to solve for 
dipole magnetic-field structure for passive fields.  The dipolar equilibrium 
magnetic field is then calculated according to the description in 
\cite{Gabler2013a}, closely following the work of \cite{Bocquet1995}. As in 
\cite{Gabler2013a}, we define a mean magnetic field strength 
\begin{equation}
\bar B = \frac{m}{(10{\rm km})^3}\left(\frac{10{\rm km}}{R}\right)^2,
\end{equation}
where $m$ is the magnetic dipole moment and $\bar B$ is equivalent
to the field strength estimate from spin down measurements. For the 
construction of our stratified equilibrium model we use the EoS of 
Akmal-Pandharipande-Ravenhall (APR) for the core \citep{Akmal1998} and of 
Douchin-Haensel (DH) for the crust \citep{Douchin2001}. Our reference 
neutron-star model has a mass of $1.4\mathrm{M}_\odot$, a radius of 
$R=12.26\,$km ($r=10.08\,$km)  and $\bar B\sim0.56B_\mathrm{pole}$ (the mean 
magnetic field strength
is roughly half the dipole magnetic field strength at the pole). 
\citet{Douchin2001} also provide the proton 
fraction $X_c^0(r)$. We take the effective masses from \cite{Chamel2008} that 
can be used to calculate the reference entrainment factor 
$\varepsilon_\star^0(r)$. Inside the inner crust, neutrons are superfluid but 
entrained strongly by the Coulomb lattice due to Bragg 
reflection~\citep{Chamel2012}. As in \cite{Gabler2016} we assume $X_c=0.9$ and 
$\varepsilon_\star=1.0$ in the inner crust and $X_c=1.0$ and 
$\varepsilon_\star=1.0$ for densities below the neutron drip point where all 
nucleons are inside the ions of the Coulomb lattice.
At the core-crust interface, the solid crust probably transitions into the 
fluid core through so called pasta phases. \cite{Passamonti2016} showed that 
these structures do not effect global magneto-elastic oscillations 
significantly, and due to yet missing consistent calculation of shear moduli 
of the pasta phases in the literature, we, thus, neglect the presence of the 
latter.
%
%=============================================================================
\section{QPO properties}\label{sec_model}

We start our analysis by collecting information of the QPOs that will be used in 
Section\,\ref{sec_constraints} to constrain the neutron star properties.
%...................................................................................................................................................................
\subsection{Properties of low-frequency QPOs}
\label{sec_lowest}

We focus on constant-phase (discrete) magneto-elastic QPOS, denoted as 
${}^l U_n$ in \cite{Gabler2016} (where $l$ as the number of maxima inside the 
crust across magnetic field lines and $n$ is the number of maxima throughout the 
star along the magnetic field lines in the region where the oscillation 
dominates). In \cite{Gabler2016} we studied the dependence of the oscillation 
frequencies of these QPOs on the magnetic-field strength and on 
$\varepsilon_\star X_c$. For a fixed value of the shear speed at the base of the 
crust, 
$c_{s,{\rm ref}}=1.3744\times 10^8$~cm/s, we obtained the following relation for the lowest-order QPO $^2U_2$:
\begin{eqnarray}
  f_{^2U_2}[\mathrm{Hz}] (c_{s, {\rm ref}})&=& 2.8\times(\varepsilon_\star 
X_c)^{-0.55} \nonumber \\
  &+& 0.66 \left(\varepsilon_\star X_c\right)^{-0.33}\bar B 
[10^{14}\mathrm{G}] 
\,. \label{eq:f2u2fit}
\end{eqnarray}
This expression is a fit to the results of the numerical simulations for values of $\varepsilon_\star X_c \in [0.046,0.695]$ and 
${\bar B} \in [7.5,20] \times10^{14}~G$.

To determine the dependence of $f_{^2U_2}$ on the shear speed $c_s$ it is not necessary to perform additional simulations. The magneto-elastic equations given in Eq.~(\ref{conservationlaw}), are invariant under the transformations
\begin{eqnarray}
t' &=&  t \, \sqrt{\frac{c_s}{c_s'}} \,,\label{eq:rescale1} \\
B'^i & = & B^i \,\sqrt{\frac{c_s'}{c_s}} \,, \label{eq:rescale2}\\
\xi'^i & = &\xi^i \,. \label{eq:rescale3}
\end{eqnarray}
Therefore, Eq.~(\ref{eq:f2u2fit}) can be rescaled accordingly to obtain the 
dependence on the value of $c_s$ as
\begin{eqnarray}
 f_{^2U_2}[\mathrm{Hz}] &=& 2.8\times(\varepsilon_\star X_c)^{-0.55}  
\sqrt{\frac{c_s}{c_{s, {\rm ref}}}} \nonumber\\ 
&+& 0.66 \times(\varepsilon_\star X_c)^{-0.33}\bar B 
[10^{14}\mathrm{G}]\,. \label{eq:f2u2_tmp}
%\nonumber\\
\end{eqnarray}
This expression has been obtained for a given neutron-star mass and EoS. Any 
change in mass or radius, either by changing the mass of the star or the EoS, 
modifies the QPO frequency. \cite{Sotani2008} showed that the frequency of 
Upper Alfv\'en QPOs $U_n$ satisfies accurate empirical relations that depend on 
the compactness of the star, $M/R$ (up to ${\mathcal O}(M/R)^2$). Note
that, even when considering the most extreme EoS and masses, the neutron star 
compactness is limited  to $M/R \sim 0.1-0.3$. Within this range, the $U_n$ 
QPO frequency varies by 50$\%$ at most~\citep{Sotani2008}. To take into 
account the dependence of frequency of the constant-phase $^2U_2$ QPOs on 
compactness, we assume that it follows the same trend as for the Upper Alfv\'en 
QPOs $U_n$ and apply the correction of \cite{Sotani2008}, which leads to
\begin{eqnarray}
  f_{^2U_2}(M/R)  =  2.5 f^{\rm ref}_{^2U_2}
  \left( 1 - 4.58 M/R + 6.06 (M/R)^2  \right)\,, \label{eq:f2u2}
\end{eqnarray}
where $f^{\rm ref}_{^2U_2}$ is the value given by Eq.~(\ref{eq:f2u2_tmp}) and 
the normalization constant is chosen so as to recover Eq.~(\ref{eq:f2u2_tmp}) 
for the particular compactness of our model ($M/R = 0.1723$).

The uncertainty in the estimation of the frequency given by Eq.~(\ref{eq:f2u2}) includes several uncertainties in the derivation of this expression. On the one hand, the error in the fit of the numerical data reported by \cite{Gabler2016} has a relative rms deviation of  $9\%$ and an absolute rms deviation of $2.65$ Hz. Additionally, there is an uncertainty of about $4\%$ due to the limited frequency resolution of the Fourier analysis done by \cite{Gabler2016}. We can thus estimate the total systematic 
error in the computation of the frequency of $^2U_2$ as $\sigma_{\rm sys}/f = 
\sqrt{0.09^2+0.04^2}\approx 0.1$.

%...................................................................................................................................................................
\subsection{Breakout and maximum magnetic-field strength}
\label{sec_breakout}

\begin{figure*}
\includegraphics[width=.9\textwidth]{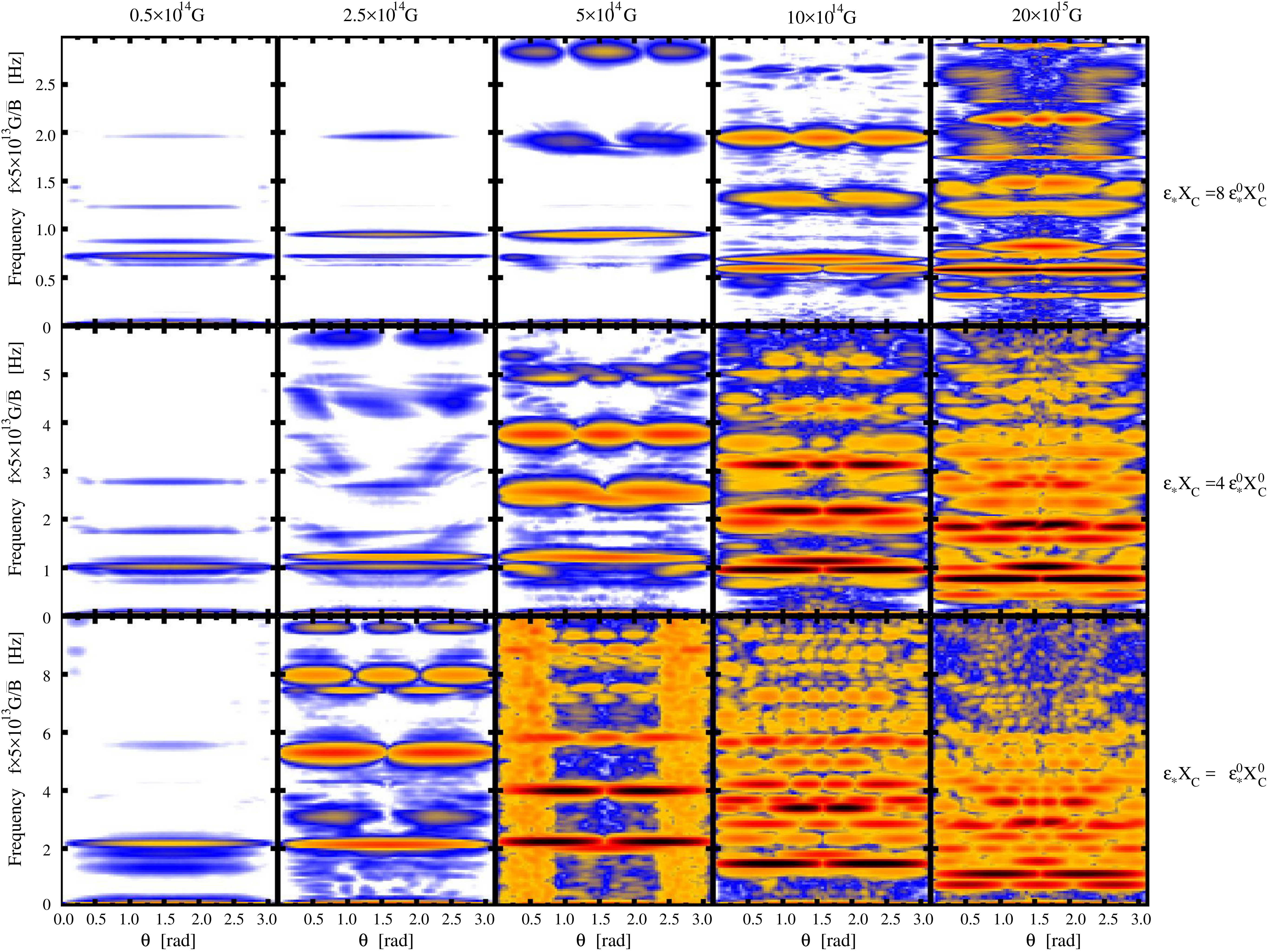}
\caption{Rescaled Fourier amplitude of the velocity close to the surface as a 
function of frequency $f$ and polar angle $\theta$ for different combinations 
of 
$\bar B$ and $\varepsilon_\star X_c$. The frequency is rescaled by the 
corresponding magnetic field as $\hat f = f\times(5\times10^{13}\,\mathrm{G} / \bar B)$. 
The color scale is logarithmic and ranges from white-blue
(minimum) to red-black (maximum)}
\label{fig_FFT_surf}
\end{figure*}

Depending on the magnetic-field strength, magnetar oscillations may be confined 
to the core or reach the surface, as shown by~\cite{Gabler2012}. In that work, 
we 
investigated the breakout to the surface of those QPOs which are confined to 
the core below a certain threshold magnetic-field strength. For a normal fluid 
core and a dipolar-like magnetic-field configuration, like the one we use here 
in our new models, the threshold for the breakout is about $\bar B
\gtrsim2\times10^{15}\,$G. We define the breakout to happen when the amplitude 
of 
the $FFT$ of $^2U_2$ is larger than that of $L_2$, the Lower ocillation which 
is confined to the closed field lines of the core.  For models including 
superfluid components, the magnetic-field strength threshold also depends on the 
proton fraction and on the entrainment, $\varepsilon_\star X_c$. 

To study the dependence of the breakout magnetic field on 
$\varepsilon_\star X_c$ we perform simulations of our new model 
series N of our reference neutron star model with different 
magnetic-field strengths and different values of $\varepsilon_\star X_c$.
We then Fourier analyze the velocity perturbation 
just below the surface of the neutron star ($r=9.74\,$km, 
$\rho=1.83\times10^8\,$g$/$cm$^3$) for different polar angles $\theta$.
The resolution of the numerical simulations is $150\times80$ zones for 
$r\times\theta=[0\,\text{km},10\,\text{km}]\times[0,\pi]$. All models have 
different Alfv\'en crossing times and thus the oscillations can have very 
different frequencies. Since, according to (\ref{eq:f2u2_tmp}), the 
frequencies depend linearly on $\bar B$ and approximately on 
$\sim 1/\sqrt{\varepsilon_\star X_c}$  \citep{Gabler2016}, we  scale the total  
simulation time 
$t$ and the time step $\Delta t$ of each model by $t=t_0 \times 
\sqrt{\varepsilon_\star X_c} / \bar B$, in order to directly 
compare the Fourier amplitude for the different models. We further rescale the obtained Fourier 
amplitude by the maximum Fourier amplitude of the Lower oscillation $L_2$ 
\cite[see][for a definition]{Gabler2016}, which should give an independent 
estimate of the magnitude of the oscillations for a given assumed initial 
perturbation, because Lower oscillations are 
not influenced by the interaction with the crust. 

In Fig.\,\ref{fig_FFT_surf} we plot in a logarithmic colour scale the Fourier amplitude at the surface of the star for a 
selection of values of $\bar B$ and $\varepsilon_\star X_c$. We rescale the 
frequencies to $\hat f=
f\times(5\times10^{13}\mathrm{G} / \bar B)$ in order to directly compare the plots for different magnetic-field strengths. The colour scale ranges from white-blue (minimum) to red-black (maximum). Each plot serves as an indicator
of whether the oscillations reach the surface (breakout) or remain inside the 
star. In the former case they could be observed
during a magnetar giant flare while in the latter they would not produce any 
observational signature. In the bottom row for models with $\varepsilon_\star
X_c=\varepsilon_\star^0 X_c^0$, i.e. in models with superfluid parameters 
expected from theoretical modelling \citep{Douchin2001},  the Fourier amplitude 
close to the 
surface increases rapidly for increasing magnetic-field strength (from left to 
right) until it saturates at around $\bar 
B\sim5\times10^{14}\,$G. At $\bar B=5\times10^{13}\,$G there is almost no Fourier signal, while for $\bar B\gtrsim10^{15}\,$G there are many different oscillations that can be identified by their angular dependence~\citep[see also 
figures 5 and 6 in][]{Gabler2016}. The corresponding frequencies at $\bar 
B=2\times10^{15}\,$G are given in Table\,\ref{tab_f_2_15} (we identify up to 
12 different QPOs - note that their relative amplitudes depend on the ad hoc 
initial perturbation.).
\begin{table}
\centering
\begin{tabular}{c | c c }
 Oscillation&$f\,$[Hz] &$\hat f = f\times(5\times10^{13}\mathrm{G}/\bar B)\,$ 
[Hz] \\
\hline 
$^3U_1$&$30$ &$0.75$\\
$^2U_2$&$44$ & $1.10$\\
$^5U_1$&$62$ & $1.55$\\ 
$^4U_2$&$71$ &$1.80$\\
$^6U_2$&$80$ &$2.00$\\
$^3U_3$&$98$ &$2.45$\\
$^5U_3$&$110$ &$2.75$\\
$^2U_4$&$115$ &$2.88$\\
$^4U_4$&$143$ &$3.48$\\
$^6U_4$&$160$ &$4.00$\\
$^7U_5$&$172$ &$4.30$\\
$^5U_5$&$195$ &$4.88$
\end{tabular}
\caption{Frequencies $f$ and rescaled frequencies $\hat f$ of our fiducial model at $\bar 
B=2\times10^{15}\,$G and $\varepsilon_\star X_c=\varepsilon_\star^0 X_c^0$.}
\label{tab_f_2_15}
\end{table}
For $\varepsilon_\star X_c=\varepsilon_\star^0 X_c^0$ we see that the 
oscillations reach 
the surface with the highest amplitudes for the first time for values of
$\bar B_\mathrm{breakout} 
  \in [2.5\times10^{14}\,$G$,5\times10^{14}\,$G$]$ (see middle panel of the 
bottom row in Fig.\,\ref{fig_FFT_surf}). The maximum amplitude does not increase 
significantly by further increasing the magnetic-field strength.

The results for two additional values of $\varepsilon_\star X_c$, namely 
$8\,\varepsilon_\star^0 X_c^0$ and $4\,\varepsilon_\star^0 X_c^0$ are shown in 
the top and middle row of panels in Fig.\,\ref{fig_FFT_surf}, respectively. In 
these models the 
coupling between the charged components and the neutrons is larger than in the 
models with $\varepsilon_\star X_c=\varepsilon_\star^0 X_c^0$. By increasing 
$\varepsilon_\star X_c$ the threshold for the breakout of the oscillations 
through 
the crust shifts to higher magnetic-field values. We performed simulations of 
all possible combinations of $\varepsilon_\star
X_c=\{1,2,4,6,8,10,15,21.7\}\times\varepsilon_\star^0X_c^0$ and $\bar 
B=\{0.5,1,2.5,5,7.5,10,15,20\}\times10^{14}\,$G in order 
to identify the corresponding value of the breakout magnetic field. Our results 
are reported in Table~\ref{tab_bbreak}.

\begin{table}
{
\setlength{\tabcolsep}{5pt}
\begin{tabular}{c c c c c c c c c c}
 $\varepsilon_\star X_c /\varepsilon_\star^0 X_c^0 $& 
%0.046&0.092&0.184&0.276&0.368&0.46&0.69&1.0\\\hline
1&2&4&6&8&10&15&21.7\\\hline
$\bar B_{\rm breakout}$  
&\multirow{2}{*}{3.75}&\multirow{2}{*}{6.25}&\multirow{2}{*}{
6.25}&\multirow{2}{*}{8.75}&\multirow{2}{*}{10}&\multirow{2}{*}{15}&\multirow{2}
{*}{15}&\multirow{2}{*}{15}\\

[$10^{14}$~G]&&&&&&&
\end{tabular}
}
\caption{Breakout magnetic field of the oscillations for different superfluid 
parameters. The breakout is defined as the point when $|FFT|$ of $^2U_2$ is 
larger than that of $L_2$. When this value is reached between two 
simulations with different magnetic field strengths we take the mean between 
the two strengths.}
\label{tab_bbreak}

\end{table}

The rescaled frequency of the identified QPOs decreases for 
$\bar B \gtrsim 5\times10^{14}\,$G in Fig.\,\ref{fig_FFT_surf}. This decrease 
is related to the offset in the frequency of the constant-phase QPOs ${}^l U_n$ 
due to the interaction with the crust, as discussed in \cite{Gabler2016}. Based 
on the results of Table~\ref{tab_bbreak}, we find the following dependence 
of the breakout field on 
$\varepsilon_\star X_c$:
\begin{equation}
\bar B_\mathrm{breakout} = 17.23 \times 10^{14}\,\mathrm{G}~ 
\sqrt{\varepsilon_\star X_c} \sqrt{\frac{c_s}{c_{s,{\rm ref}}}} . 
\label{eq:bbreak}
\end{equation}
Note that the dependence on $c_s$ is obtained using the rescaling of 
Eqs.~(\ref{eq:rescale1})-(\ref{eq:rescale3}), as described in the previous 
section. 
The relative rms deviation of this fit with respect to the data in 
Table~\ref{tab_bbreak} is $13\%$ and the relative error in the determination of 
the $\bar B_{\rm breakout}$ itself is $33\%$, due to the limited number of 
simulations. The systematic error by using this 
expression is $\sigma_{\rm sys} / \bar B_\mathrm{breakout}  = 0.35$ (which 
should improve significantly if a larger number of simulations in the vicinity 
of $\bar B_{\rm breakout}$ is used). 
When applying Eq.\,\ref{eq:bbreak}, we implicitely assume that the 
magnetic field strength does not change significantly through the crust, and 
thus $\bar B_\mathrm{breakout}$ is not very sensitve to crust thickness and 
compactness of the particular stellar model.

While the breakout field represents the minimum field at which QPOs can reach 
the surface, there is another limitation on the {\it maximum} field too, as 
this must be compatible with the information provided by observations. If the 
Alfv\'en velocity inside the crust is much larger than the shear speed there, 
QPOs disappear and an Alfv\'en continuum appears~\citep{Gabler2013a, 
Passamonti2013}. This continuum cannot sustain long-lived QPOs and would not 
explain, in particular, the observations of high-frequency 
QPOs~\citep{Levin2006,Levin2007}. Therefore, the maximum magnetic field at the 
crust, defined as the value at which Alfv\'en and shear speed are equal, is
\begin{equation}
{\bar B}_\mathrm{max} =  1.79\, c_s \sqrt{\rho_{\rm cc}}\,, \label{eq:bmax}
\end{equation}
where the density at the core-crust interface in our model is $\rho_{\rm 
cc}=1.108\times10^{14}$\,g/cm$^3$, and we have taken into account that the 
surface field at the pole is $1.79$ times larger than $\bar B$, for our 
reference magnetic-field structure (see Section~\ref{sec_theory}). This maximum 
field is a rough estimate of the typical magnetic fields at which the transition 
to an Alfv\'en-dominated spectrum appears. From \cite{Gabler2013a} we estimate 
that this transition occurs at magnetic fields deviating no more than $50\%$ of 
this maximum field. Therefore, the systematic error in the expression for the 
maximum field is $\sigma_{\rm sys}/{\bar B}_\mathrm{max} =0.5$.

%............................................................................
\subsection{High-frequency QPOs}\label{sec_high}

In the preceding sections \citep[and in][]{Gabler2016} we focused on 
low-frequency QPOs  with frequencies up to $155\,$Hz. These QPOs can be 
explained with the fundamental magneto-elastic oscillations and some lower 
overtones. To identify the high-frequency QPOs at $0.625\,$kHz and $1.84\,$kHz 
with the same kind of oscillations one needs a high overtone of $n\sim30$ or 
$90$, 
respectively. However, to us it seems hard to explain why only these 
particular overtones should be excited and not many more. A possible explanation 
was first presented in \cite{Gabler2013a} and was confirmed by 
\cite{Passamonti2013}. Both groups find a preferential 
excitation of only a few particular high Alfv\'en overtones in the core: Let the 
crust be initially perturbed by a deformation with spatial 
structure corresponding to the eigenfunction of a crustal shear mode 
(as may happen during a giant flare). The simulations then show that  
such an initial perturbation preferentially excites one or a 
few dominant global magneto-elastic oscillations at different frequencies for 
dipolar fields $\sim 10^{15}$G, only if the core is assumed to be superfluid. In 
the latter case the shear terms in the crust are not yet completely dominated by 
the Alfv\'en terms.

In the following, we will refer to  crustal shear modes of 
 \textit{non-magnetized} models as \textit{pure} crustal shear modes and
denote them as ${}^l t^{(0)}_n$ (the index $(0)$ indicating that they are
defined in the limit of vanishing magnetic field). Although pure 
crustal shear modes do not exist for such magnetic field strengths of order 
$\sim 10^{15}$G, one may regard (in simplified terms) the existence of the 
preferentially excited global magneto-elastic QPOs as a resonant excitation 
between a \textit{magnetically modified} shear oscillation in the crust and a 
torsional Alfv\'en oscillation in the core.

Which particular overtones of the \textit{magnetically modified} shear 
oscillation are excited depends on the spatial 
structure of the applied perturbation. 
To resolve the much finer spatial structures in radial direction of these 
oscillations we need to increase the resolution of our simulations. Therefore, 
we have performed simulations of the models of series N with $(300\times100)$, 
$(400\times132)$, and $(500\times167)$ zones for 
$r\times\theta=[0\mathrm{km},10\mathrm{km}]\times[0,\pi/2]$. All simulations 
lead to qualitatively similar results. For the results discussed in this section 
we assume $\varepsilon_\star X_c=\varepsilon_\star^0 X_c^0$.

\begin{figure}
\includegraphics[width=.47\textwidth]{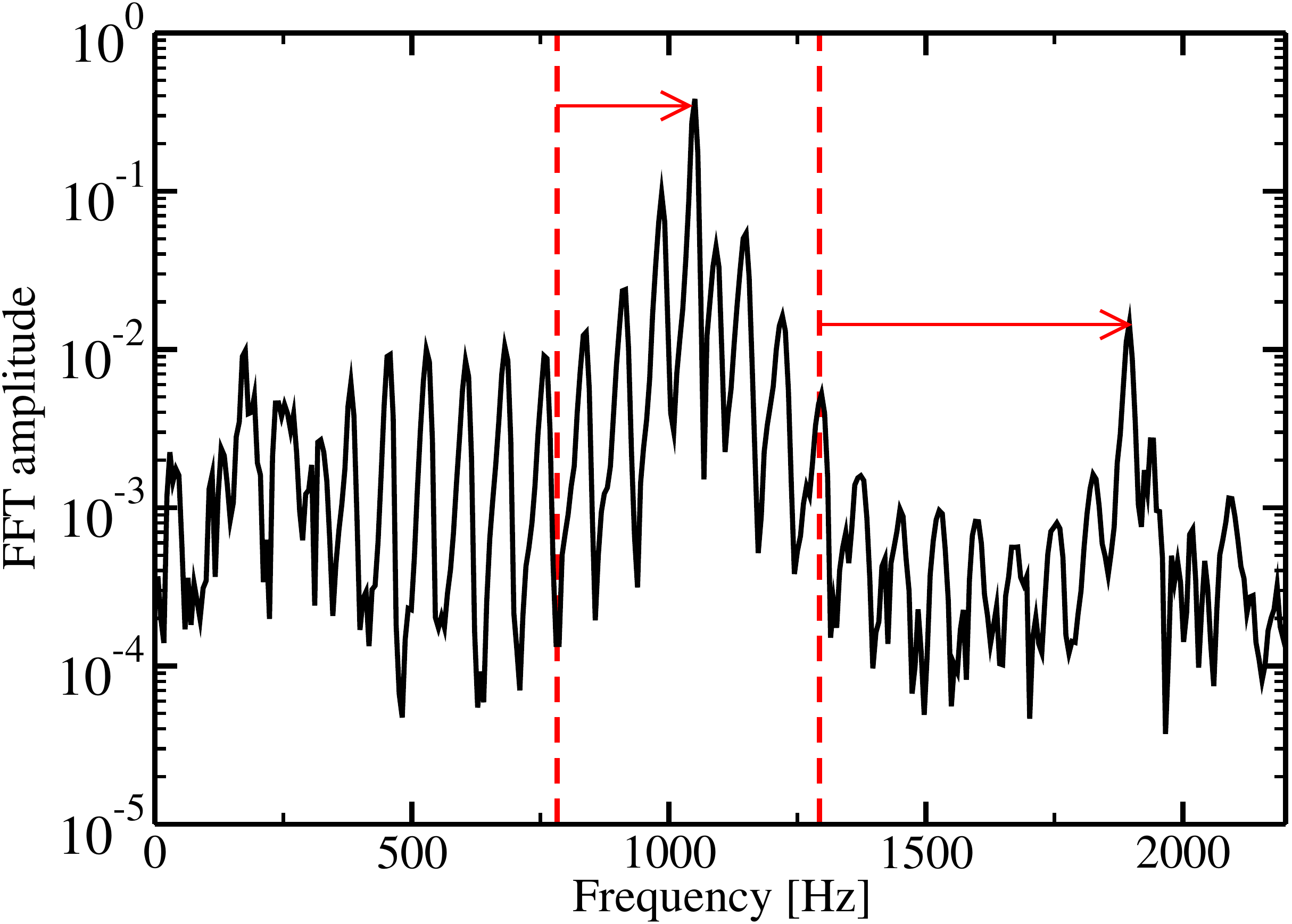}
\caption{Fourier amplitude of the velocity at one point inside the crust 
($r=9.4\,$km, $\theta=0.12$). The initial perturbation uses the eigenfunction 
of a pure $^2t^{(0)}_1$ crustal 
shear mode. Dashed red lines indicate the frequencies of the two radial 
overtones of the pure crustal shear modes, $f_{^2t^{(0)}_1}=0.79\,$kHz and 
$f_{^2t_2}=1.29\,$kHz. The strongest oscillations during the simulation have 
significantly higher frequencies $f\sim1.05\,$kHz and $f\sim1.9\,$kHz, as 
indicated by the arrows. }
\label{fig_n1_FFT}
\end{figure}

\begin{figure}
\includegraphics[width=.47\textwidth]{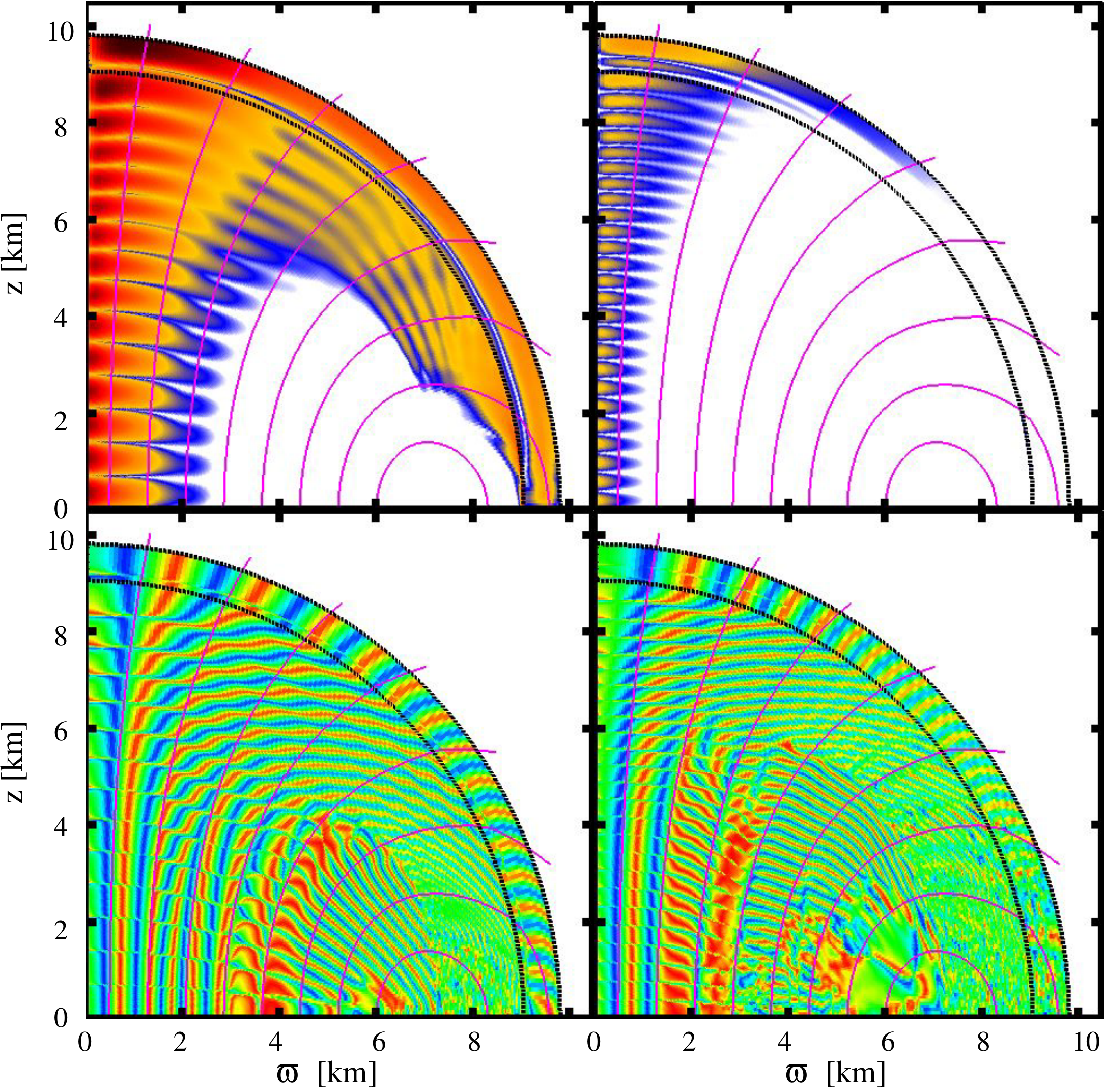}
\caption{Structure of the Fourier transform and phase of two strongly excited magneto-elastic oscillations $^2U_{30}$ and $^2U_{54}$ with $f=1.05\,$kHz (left panels) and $f=1.9\,$kHz (right panels), respectively. Magenta lines indicate the magnetic-field lines and black lines indicate the location of the crust. Top: Fourier amplitude. The colour scale is logarithmic and ranges 
from white-blue (minimum) to red-black (maximum). Bottom: Phase. The colour scale is linear and ranges from blue ($-\pi/2$) to orange-red ($\pi/2$).}
\label{fig_n1_structure}
\end{figure}

   We initiate the simulations with an initial perturbation in the form of the eigenfunction of a
$^2t^{(0)}_1$ pure crustal shear mode. Fig.\,\ref{fig_n1_FFT} plots, for the 
highest resolution simulation, the Fourier amplitude at one point inside the 
crust ($r=9.4\,$km, $\theta=0.12$) for an evolution time of $160\,$ms and $\bar 
B=10^{15}\,$G. There are two local maxima at $f\sim1.05\,$kHz and 
$f\sim1.9\,$kHz, respectively. The global structure and the phase of the Fourier 
transform of these oscillations are plotted in Fig.\,\ref{fig_n1_structure}. 
These oscillations correspond to the Upper magneto-elastic QPOs 
$^2U_{30}$ and $^2U_{54}$, in the notation of \citep{Gabler2016}. The structure 
inside the crust along the radial direction is very similar to the $^lt^{(0)}_1$ 
and $^lt^{(0)}_2$ pure crustal shear modes. We note that there are 
either one or two radial nodes roughly at the radius where the two pure
crustal shear modes have nodes. In the resonance picture, the two 
excited QPOs could thus be regarded as resonances between two torsional Alfv\'en 
Upper QPOs in the core (with $28$ and $50$ maxima in the core region) and  
magnetically modified crustal shear modes in the crust, which we denote as 
$^lt_1$ and $^lt_2$, respectively, which result in the global magneto-elastic 
QPOs $^2U_{30}$ and 
$^2U_{54}$. 

The phase of these oscillations is continuous everywhere, also in the crust, as 
is shown in Fig.\,\ref{fig_n1_structure}. 
This is in contrast to the constant phase for low-frequency QPOs in 
superfluid models and could be 
an artefact of the limited resolution of the simulations and/or a consequence of 
perturbing the star with a crustal shear mode rather than directly exciting the 
oscillation itself. We note that when using lower resolution we do not find a 
constant phase after several mode-recycling steps, i.e.~after exciting the star 
always with the  oscillation pattern of the previous simulation at a given 
frequency. However, it is not surprising that this oscillation has a continuous 
phase, because the $n\geq1$ crustal shear modes travel in the radial direction, 
which coincides with the direction of the magnetic field lines close to the 
poles. Therefore, the crust hardly couples the field lines together in the 
perpendicular direction and no constant-phase oscillations form. This is in 
contrast to the constant-phase oscillations at lower frequencies described in 
\cite{Gabler2013a,Gabler2016}.

Note that, in the Fourier transform of Fig.\,\ref{fig_n1_FFT} many 
more QPOs than just $^2U_{30}$ and $^2U_{54}$ are excited. These are 
all overtones $^2U_n$ with different number $n$ of maxima along the magnetic 
field lines. However, the $^2U_{30}$ and $^2U_{54}$ QPOs form distinct 
local maxima in the Fourier spectrum. note that $^2U_{54}$ has a lower 
amplitude than some QPOs asound $^2U_{30}$, because the initial excitation 
in form of $f_{^2t^{(0)}_2}$ preferentially excites QPOs with only one node 
inside the crust. The 
red dashed lines in Fig.\,\ref{fig_n1_FFT} indicate the frequencies of 
what would be the pure crustal shear modes $f_{^2t^{(0)}_1}=0.79\,$kHz 
and $f_{^2t^{(0)}_2}=1.29\,$kHz in a nonmagnetized model, 
respectively. Clearly, at $f=0.79\,$kHz there is no oscillation, while the 
$f=1.29\,$kHz frequency happens to (coincidentally) agree with one of 
the many peaks in the Furier spectrum.  Moreover, the spatial structure of the 
$f=1.29\,$kHz peak has only one node inside the crust in the radial direction 
and, thus, can not be related directly to ${^2t^{(0)}_2}$.

In view of the above findings, the observed high-frequency QPOs at 
$0.625$\,kHz and $1.84\,$kHz in the giant flare of SGR 1806-20 cannot 
be interpreted as corresponding to pure crustal shear modes.
Instead, they may be due to preferential resonant excitation of particular, 
global magneto-elastic QPOs, due to the actual seismic event associated with the 
generation of giant flares. In the following, we will assume that this 
preferential excitation happens at two different resonant frequencies ${}^2 U_n$ 
associated with what can be regarded as the two lowest-order magnetically 
modified crustal shear modes  $^2t_1$ and $^2t_2$. Because the order $n$ of the 
resonantly excited QPOs depend on the EoS, mass and magnetic 
field, and to simplify the discussion, we will from now on refer to the 
${}^lt_1$ and 
$^lt_2$ QPOs, meaning, in all cases, the corresponding resonantly excited 
magnetoelastic, high-frequency  ${}^2
U_n$. Our reference to the  $^2t_1$ and $^2t_2$ frequencies should thus not be 
interpreted as meaning pure crustal modes). 

\begin{figure}
\includegraphics[width=.47\textwidth]{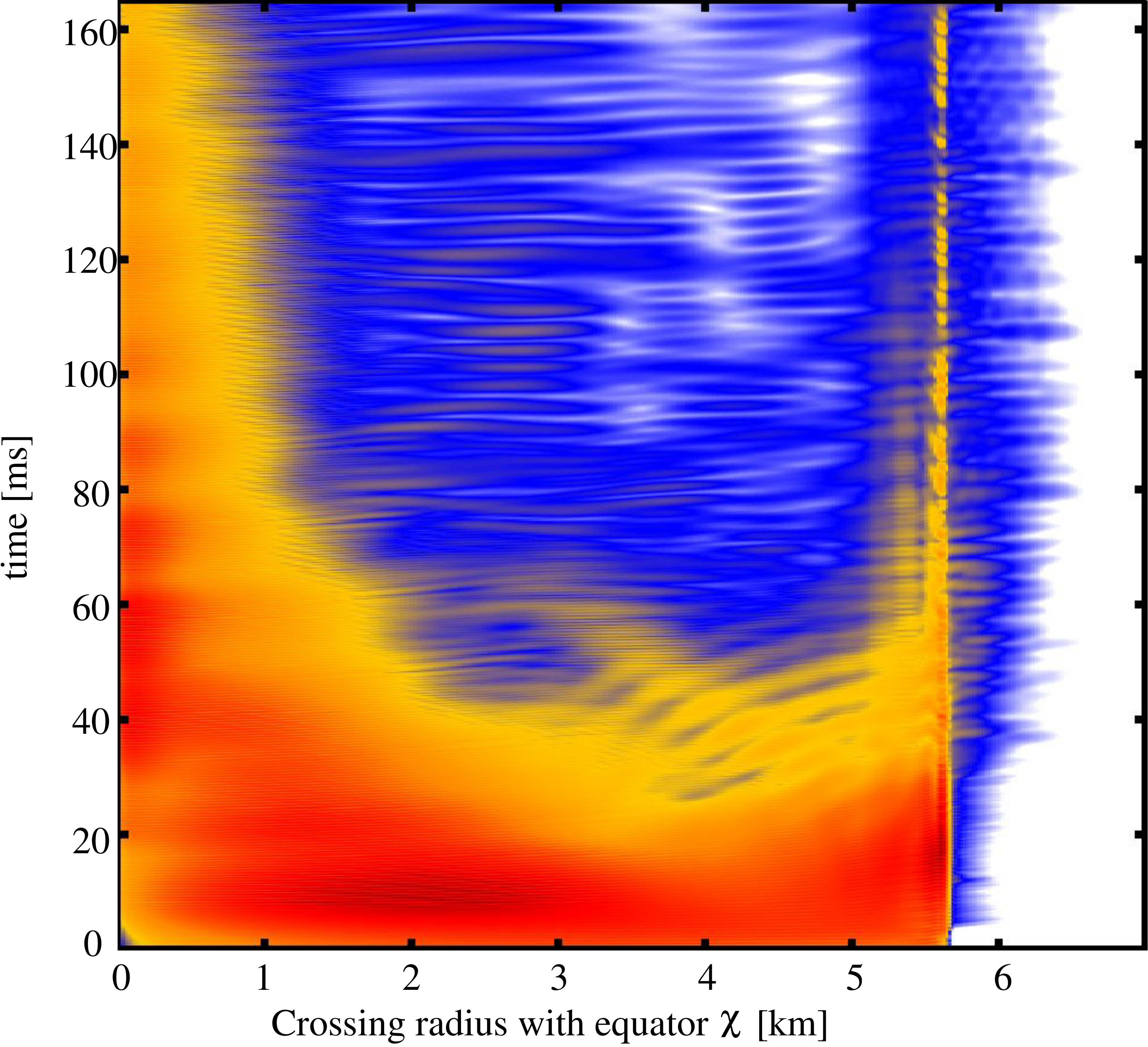}
\caption{Rescaled kinetic plus magnetic energy along field lines as a function 
of time for a simulation with $\bar B=10^{15}\,$G. Each field line is labeled 
by its crossing radius with the equatorial 
plane $\chi$. The closed field line region starts at $\chi\sim5.5\,$km. The 
color scale is logarithmic and ranges from white-blue (minimum) to 
red-black (maximum).}
\label{fig_n1_ener}
\end{figure}
\begin{table*}
\begin{tabular}{c | l c c c c c c c c c}
 $\bar B[10^{14}\,{\rm G}]$
&&$0.0$&$0.1$&$0.5$&$1.0$&$2.5$&$5.0$&$7.5$&$10$&$20$\\\hline
 $f_{^2t_1}$ [kHz]&$\theta=0$&&&&&&$0.87$&$0.96$&$1.05$&$1.48$\\
 &$\theta=\pi/4$&$0.80$&$0.77$&$0.77$&$0.78$\\
 &$\theta=\pi/2$&&&&&$0.78$&$0.79$&$0.79$&$0.81$&$0.87$\\\hline
 $f_{^2t_2}$ [kHz] &$\theta=0$&&&&&&($1.51$)&($1.71)$&$(1.85)$&$(>2.0)$\\
 &$\theta=\pi/4$&$1.29$&$1.26$&$1.27$&$1.28$&$1.34$&$(1.46)$&\\
 &$\theta=\pi/2$&&&&&$1.26$&$(1.29)$&$(1.30)$&$(1.33)$\\\hline
 %$f_{p2}$ [Hz]&&&&&&$893$&$990$&$1093$
\end{tabular}
\caption{Oscillation frequencies at local maxima of the Fourier amplitude peaks. 
The values in parenthesis indicate that their determination is not accurate due 
to the limited resolution of the numerical simulation. The frequencies for $\bar 
B=0.0\,$G are obtained by an eigenmode calculation and may thus differ slightly 
from those obtained by analyzing the Fourier amplitude of the evolutions. We 
name the oscillations at $\theta=\pi/2$ as crustal shear modes and the ones at 
$\theta=0$ $^2U_n$. The latter have different number of maxima $n$ inside the 
core for different magnetic-field strength $\bar B$.}
\label{tab_n1}
\end{table*}

Next, we study the behaviour of the high-frequency QPOs with varying 
magnetic-field strength. To save computational power and time we use a medium
resolution of $400\times132$. With this grid we obtain the following eigenfrequencies 
for the pure crustal shear modes $^2t^{(0)}_{1}$, $^2t^{(0)}_{2}$ and $^2t^{(0)}_{3}$ in the absence of a magnetic field: $f_{^2t^{(0)}_1}=0.80\,$Hz, 
$f_{^2t^{(0)}_2}=1.29\,$kHz and $f_{^2t^{(0)}_3}=1.8\,$kHz, respectively. For 
non-vanishing magnetic fields, the oscillations of the crust are absorbed 
rapidly into the core, 
even at relatively low magnetic field strengths ($\bar B 
\lesssim2.5\times10^{14}$G), but there is sufficient time to determine their 
frequencies. For very weak magnetic fields we expect to recover the 
frequencies of the 
eigenmode calculation approximately, and indeed the results of the simulation are off by only a few per cent at $\bar 
B=10^{13}\,$G: $f_{^2t_1}^\mathrm{}(10^{13}\text{G})=0.77\,$Hz and 
$f_{^2t_2}^\mathrm{}(10^{13}\text{G})=1.26\,$kHz (see Table\,\ref{tab_n1}).
For magnetic-field strengths up to $\bar B\leq10^{14}\,G$ the oscillation 
structure does not change significantly and the frequencies only increase 
slightly $f_{^2t_1}^\mathrm{}(10^{14}\text{G})=0.78\,$Hz and 
$f_{^2t_2}^\mathrm{}(10^{14}\text{G})=1.28\,$kHz. However, the crustal oscillations are damped much faster with increasing magnetic field strength. 

For $\bar B> 10^{14}\,$G the angular oscillation pattern splits into two 
branches, during the time evolution.
One branch  stays confined to the crust and its maximum 
amplitude shifts towards the equator with increasing field strength \citep[see 
also Figure 4 in][for a similar oscillation pattern obtained in the case of a 
normal fluid core]{Gabler2013b}. These localized oscillations are 
dominated by the shear 
inside the crust. In the equatorial region the magnetic field is 
approximately parallel to the $\theta$-direction and, hence, an oscillation 
that propagates preferentially along the radial direction like the $n\geqslant1$ 
crustal shear modes is not prone to interact with the magnetic 
field. This kind of shear-like oscillations are rapidly damped in regions where the 
magnetic field has a significant radial component. The frequency of this 
localized oscillation just outside  the region of closed field lines only 
increase mildly with the magnetic-field strength: e.g. it becomes 0.81 
kHz and 1.33 kHz (for oscillations that can be associated with 
$^2t^{(0)}_{1}$ and $^2t^{(0)}_{2}$,  respectively at $10^{15}$G.
The other branch of oscillations are global magneto-elastic 
oscillations $^2U_n$ that are concentrated around the polar axis (see
Fig.\,\ref{fig_n1_structure}). 
These oscillations appear for $\bar 
B\gtrsim5\times10^{14}\,$G and their frequencies increase strongly with 
increasing magnetic-field strength, e.g. 
$f_{^2t_1}^\mathrm{}(2\times10^{15}\text{G})=1.48\,$kHz and 
$f_{^2t_2}^\mathrm{}(10^{15}\text{G})=1.85\,$kHz, respectively (see Table\,\ref{tab_n1})
.  

To illustrate the different time dependence of the two branches we plot in 
Fig.\,\ref{fig_n1_ener} the rescaled kinetic plus magnetic energy of the 
oscillations for each magnetic field line as a function 
of time.  The horizontal axis in this plot is the radius $\chi$ at which the 
magnetic field lines cross the equator \citep{Cerda2008}. The open magnetic 
field lines reach up to $\chi\sim5.5\,$km. Initially, we apply a 
perturbation corresponding to the pure 
crustal shear mode $^2t^{(0)}_1$ and the strongest excitation is seen along the 
field lines at $\chi\sim2\,$km. These lines enter the crust at $\theta=\pi/4$ 
and, as the figure shows, the initial excitation of the crustal shear mode has its 
maximum amplitude around this angle. After $20\,$ms the energy redistributes 
and the major part moves to the pole $\chi=0$, while some part moves to the 
field lines entering the crust close to the equator, $\chi\lesssim5.5\,$km.
At $t>100\,$ms the oscillations near the equator almost completely disappear.
Only minimal energy leaks into the region of closed field lines at 
$\chi\gtrsim5.5\,$km due to a numerical effect caused by the 
steep gradients at the core-crust boundary.

The damping time $\tau$ of the localized shear oscillations just outside the 
region of closed field lines is about 
$\tau\sim10\,$ms for the model in Fig. \ref{fig_n1_ener} and thus comparable 
to what was found by 
\cite{vanHoven2012}. Such a short damping time can not account for the 
observation of the $625\,$Hz QPO in SGR 1806-20, as its damping time derived 
from observations 
is longer than at least about $500\,$ms \citep{Huppenkothen2014c}.
On the other hand, the high-frequency
magneto-elastic QPO close to the pole survives for much longer 
times. Its damping time, obtained from the simulation, is limited by the numerical damping of our scheme 
\citep[see also][]{Gabler2012}. The kinetic plus magnetic energy in all 
oscillations over the whole neutron star 
volume decreases at the same rate $\tau\sim50\,$ms as the energy in the field 
lines close to pole that mainly represent the excited high-frequency 
magneto-elastic QPO. Therefore, a damping time of the observations of about 
$500\,$ms could still be consistent with the simulations for this kind of 
oscillations.

In \cite{Gabler2016} we showed that the low-frequency, constant-phase 
oscillations only appear in a certain range of magnetic-field strengths. For the 
high-frequency oscillations, approximately the same range holds. If the magnetic 
field is too weak or too strong compared to the shear modulus, shear 
oscillations in the crust do not resonate with torsional Alfv\'en oscillations 
in the core.  Moreover, when increasing the superfluid parameter 
$\varepsilon_\star X_c$, the range of magnetic-field strengths for observing 
constant-phase oscillations shrinks. Additionally, this range depends weakly on 
the particular way the interface between the crust and the core is treated. We 
have assumed either a discontinuous change or a smooth transition of the shear 
modulus through the pasta phase. In the discontinuous case the range is 
narrower. As in \cite{Gabler2013b}, our new simulations also show a 
high-frequency, shear-like oscillation close to the equator in all cases. 
However, as we already mentioned there, this oscillation has two problems to 
explain the observed frequencies: firstly, it is transient and rapidly damped, 
and secondly, it is limited to a region close to the equator where magnetic 
field lines barely reach the exterior of the star. Being almost confined to the 
interior these oscillations cannot easily modulate the signal during a giant 
flare. 

To compare with observations, we fit the frequencies of the resonantly excited 
high-frequency magneto-elastic QPOs given in Table\,\ref{tab_n1} (near the 
pole), as a function of $\bar B_{14}\equiv\bar B /10^{14}{\rm G}$
\begin{eqnarray}
 f_{^2t_n}&\simeq&f_{^2t^{(0)}_n} \left(1+a_{2t_n} 
\bar 
B^2_{14}\right)^{1/2}\,\label{eq:crustfreq_0}.
\end{eqnarray}
We find $a_{2t_1}=0.0063$  and $a_{2t_2}=0.012$. Note, however, 
that the fitting formula (\ref{eq:crustfreq_0}) is only valid for the 
particular magnetic-field structure used in our simulations. Irrespective of its 
structure, the effect of the magnetic field is always to increase the frequency 
of the magnetically modified crustal oscillation $^2t_n$ above that of the pure 
crustal mode, ${^2t^{(0)}_n}$:
\begin{equation}
 f_{^2t_n} \ge f_{^2t^{(0)}_n}.\label{eq:crustfreq_1}
\end{equation}

\cite{Samuelsson2007} estimated that the frequencies of pure crustal modes 
with $n\ge1$ assuming constant shear speed inside the crust are proportional to 
$f_{^2t^{(0)}_n}\sim \left(1-\frac{2M}{R}\right)\frac{2 c_s}{n\Delta 
r}$. Comparing to our previous study where we calculated crustal shear mode 
frequencies \citep{Gabler2012}, we find that the scaling with $M/R$ and $c_s$ 
roughly holds, but the absolute frequency differs significantly because 
$c_s$, which we determine at the core-crust interface, is not equivalent to the 
mean shear velocity. Thus, we prefer to use the following approximation

\begin{equation}
f_{^2t^{(0)}_n}\simeq\frac{\left(1-\frac{2M}{R}\right)\frac{c_s}{\Delta 
r}}{\left(1-\frac{2M_\mathrm{ref}}{R_\mathrm{ref}}\right)\frac{c_s^\mathrm
{ref } } { \Delta r_\mathrm{ref}}} 
f^\mathrm{ref}_{^2t^{(0)}_n}\label{eq:crustfreq}
\end{equation}
where $\Delta r$ indicates the size of the crust and $\mu_{cc}$ is the 
shear modulus at the core-crust interface. The values of our reference model 
are $M_\mathrm{ref}/R_\mathrm{ref}=0.172$ 
, $\Delta r_\mathrm{ref}=870\,$m, $f^\mathrm{ref}_{^2t^{(0)}_1}=780\,$Hz, and 
$f^\mathrm{ref}_{^2t^{(0)}_2}=1290\,$Hz. Here, we have neglected the 
dependence on $l$ ($\lesssim 0.2\%$ correction for $\leqslant 2$). The 
value of the shear speed in the crust is, however, not constant. 
We thus contrast the above formula with the calculation of crustal shear modes 
in \cite{Sotani2007} and find a maximal deviation of $35\%$, which we take as 
the corresponding systematic error $\sigma_{\rm sys} / f_{^2t^{(0)}_n} =0.35$. 

The constraint (\ref{eq:crustfreq_1}) on the frequency of resonantly excited, high-frequency QPOs depends implicitly on the compactness of the star, because the size of the crust is related to such compactness. Namely, for a given EoS the crust is larger for less compact (less massive) stars. However, since this relation is model dependent, we prefer to follow a more general approach and leave the size of the crust as a free parameter.  

So far we have presented a lower bound for  the $^2t_n$ frequencies. 
Additionally, the frequency of these QPOs, cannot become arbitrarily large with 
increasing magnetic field. The reason is that for sufficiently large magnetic 
fields, the $^2t_n$ QPOs are dissolved into the Alfv\'en continuum (see 
also the corresponding discussion for the low-frequency oscillations in 
Section\,\ref{sec_breakout}). In our new simulations (see Table~\ref{tab_n1}) 
the continuum starts 
to appear at a magnetic field just above $2\times10^{15}$~G, at which the 
frequency of the $^2t_n$ QPOs is at most a factor $1.85$ larger than the  
corresponding pure crustal shear eigenfrequency. Similar results were found by 
\cite{vanHoven2012}, i.e. an increase of about a factor $2$ in the frequency of 
the $n\ge1$ modes at $2\times10^{15}$~G. Therefore, we introduce the 
additional constraint
\begin{equation}
f_{^2t_n} < 2 f_{^2t^{(0)}_n}\,. \label{eq:highmax}
\end{equation}
To double the frequency of the ${^2t_n}$ QPOs, the magnetic field 
needs to be so strong, that the Alfv\'en continuum starts to dominate and no 
resonant oscillations are allowed.

%
%=============================================================================
\section{Constraining neutron star properties}\label{sec_constraints}

%..........................................................................................................................
\subsection{Example model reproducing observed frequencies}

In Section~\ref{sec_lowest} we  discussed the dependency of the frequency of the 
low-frequency, constant-phase $^2U_2$ mode on the different parameters of 
the system. Assuming that  the magnetic-field strength, the shear speed, and the 
compactness of the star are known, a clear observational identification of 
$^2U_2$ would allow us to constrain the parameter $\varepsilon_\star X_c$. 
As an example, let us assume that $^2U_2$ is one of the observed QPOs 
in SGR 1806-20 with the lowest frequencies, $18$, $26$ or $30$ Hz,  
with $\bar B=2\times10^{15}\,$G. For the given EoS and magnetic-field 
configuration we thus find $\varepsilon_\star X_c(r=0)=0.73$, $0.27$ or $0.19$, 
correspondingly. These values are significantly higher than 
the value  we expect from the EoS, $\varepsilon_\star^0 X^0_c(r=0)=0.046$, which 
indicates that, for this model, either the proton fraction has to be higher, or 
the entrainment should be different, or both. 

\begin{figure}
\includegraphics[width=.47\textwidth]{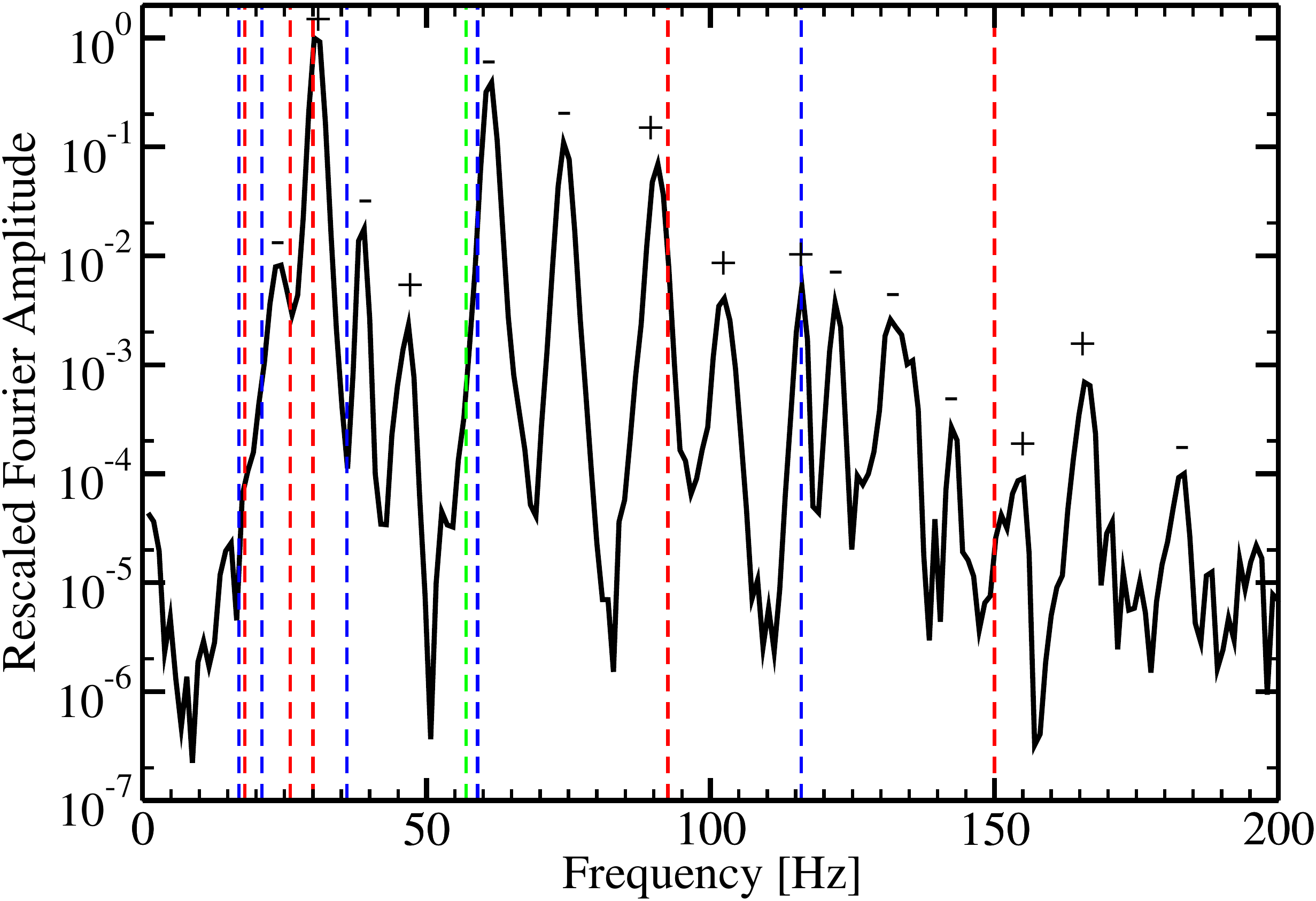}
\caption{Fourier amplitude at one point inside the crust  $r=9.4\,$km and 
$\theta=0.6$. The dashed red lines indicate the frequencies of the observed 
QPOs during the giant flare of SGR 1806-20, the blue lines show additional 
frequencies found by \protect\cite{Hambaryan2011}, and the green line shows the 
frequency found in normal bursts by \protect\cite{Huppenkothen2014b}. Plus 
(minus) signs indicate oscillations that are symmetric (antisymmetric) in 
$B_\varphi$  w.r.t. the equator.
}
\label{fig_fit_freqs}
\end{figure}

Similarly, we could constrain the magnetic field strength if we assume that the 
EoS and related quantities, i.e. $\varepsilon_\star^0 X^0_c(r=0)=0.046$, are 
known. In this case, and by using Eq.~(\ref{eq:f2u2}), we obtain a dipolar 
magnetic-field strength of $\bar B\sim1.1\times10^{15}\,$G. We
performed a new simulation for this case and find that the corresponding 
spectrum is very 
similar to that of the previous case. In Fig.\,\ref{fig_fit_freqs} we plot the Fourier amplitude at a point 
near the surface of the star. We identify the 14 strongest frequencies, which 
are displayed in Table\,\ref{tab_fit_f}. Since this model was fitted to 
reproduce 
the oscillation around $30\,$Hz this frequency is recovered. Additionally, we 
also find significant oscillations near all other higher frequencies 
($f<200\,$Hz) that have been reported for SGR 1806-20, i.e. $f\sim36, 
59, 92, 116, 150\,$Hz. Naturally, we can only explain one of the lower 
frequencies $f<30\,$Hz, because we matched the $^2U_2$ mode and only the $^3U_1$ 
is expected to have a lower frequency. However, the detection of these 
oscillation frequencies during a giant flare is likely very complicated due to 
the temporal variation of the flare itself that may produce features in the 
spectral analysis at these frequencies. 

In Fig.~\ref{fig_fit_freqs} we also show the symmetry with respect to the equator of the magnetic-field perturbation for each oscillation. As observed in \cite{Gabler2014a, Gabler2014b} only symmetric (+) oscillations are able to modulate the exterior magnetic field and, hence, the corresponding emission. In the current model, we could thus accommodate the QPOs with $30, 92, 116,$ and $150\,$Hz, while the oscillations near $36$ and $59\,$Hz should not be observed due to the antisymmetric character of the corresponding magneto-elastic oscillations.

\begin{table}
\begin{tabular}{c | c c c c c c c}
Oscillation&$^3U_1$&$^2U_2$&$^5U_1$&$^4U_2$&$^3U_3$&$^5U_3$&$^4U_4$\\
 $f\,$[Hz]&24 &30&39&47&61&74&91\\
\hline
Oscillation&$^6U_4$&$^8U
 _4$&$^5U_5$&$^7U_5$&$^9U_5$&$^6U_6$&$^8U_6$\\
 $f\,$[Hz]&102&116&122&132&142&153&169\\ 
\end{tabular}
\caption{Dominating oscillation frequencies of our fiducial model at $\bar 
B=1.1\times10^{15}\,$G and $\varepsilon_\star X_c=\varepsilon_\star^0 X_c^0$.}
\label{tab_fit_f}
\end{table}

 Although the magnetic-field configuration in magnetars is currently unknown, 
the above examples show that it would be possible to constrain properties of 
the high-density matter of neutron stars through asteroseismology, once some 
parameters such as the magnetic-field strength, magnetic-field configuration or 
the mass are constrained by other observations.

%...............................................................................
...........................................
\subsection{Combining constraints}

\begin{figure*}
\includegraphics[width=.47\textwidth]{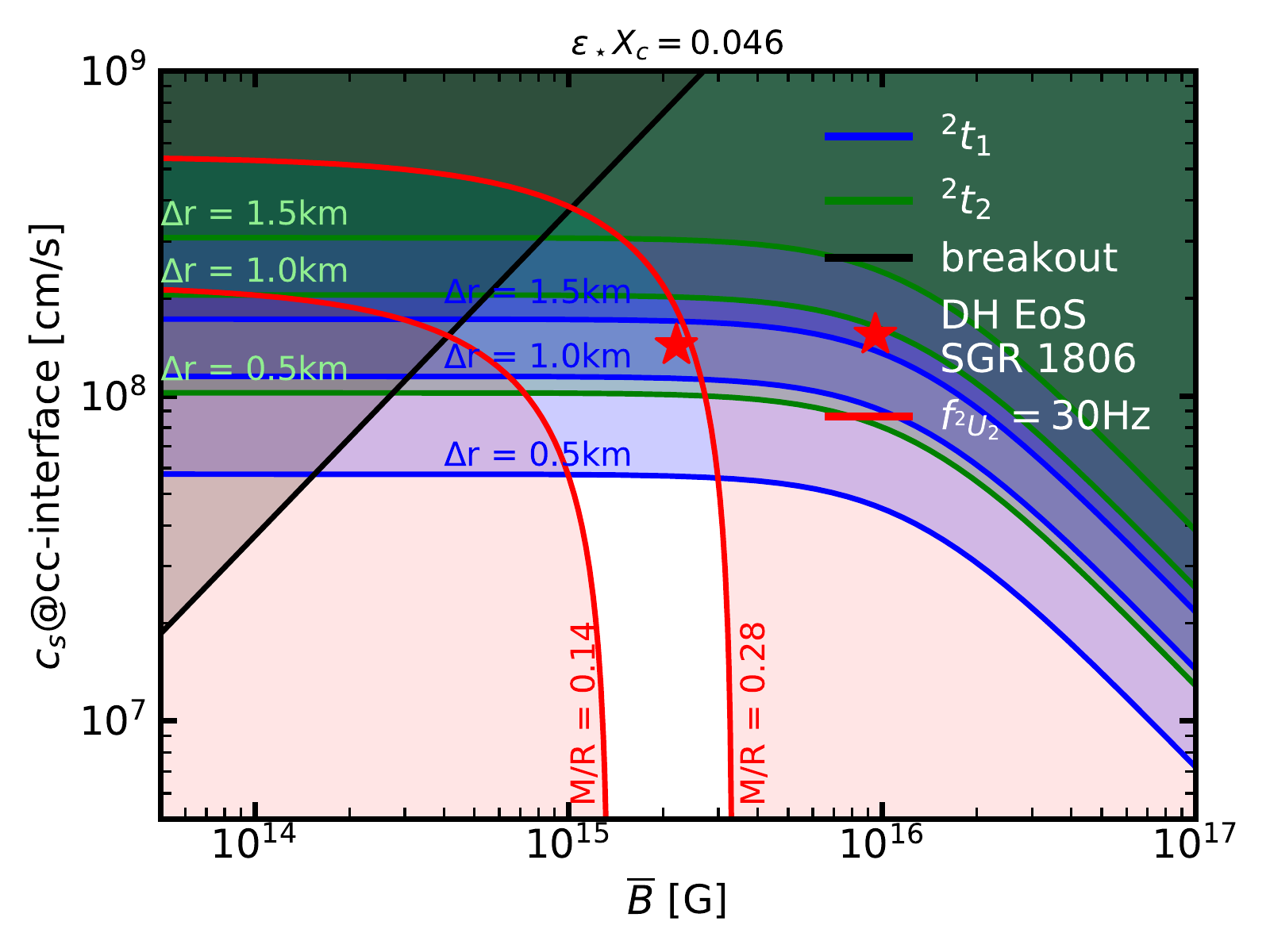}
\includegraphics[width=.47\textwidth]{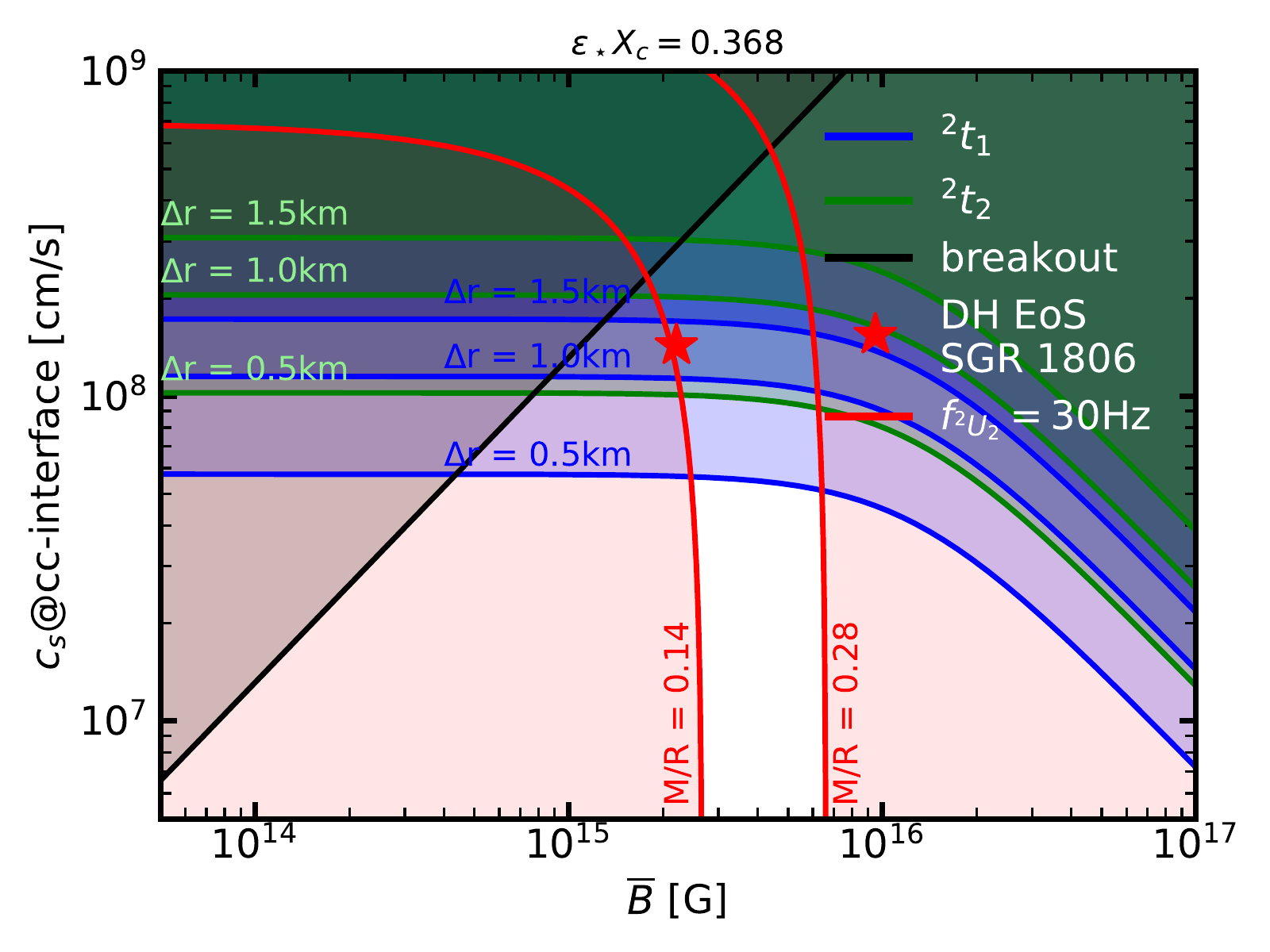}
\caption{Constraints on the $c_{s}-\bar B-$plane obtained from the 
magneto-elastic QPOs. The frequencies of the resonant high-frequency QPOs have 
to be below the observed frequencies for $^2t_1$ (blue lines) and $^2t_2$ (green 
lines). For better visibility we only show the strongest constrain, i.e. for 
$M/R=0.14$, for the high frequency QPOs.
The magnetic field has to be strong enough to break out of the crust. 
The threshold is given by the black lines. The two red lines indicate the 
frequency of $^2U_2$ matched to $30\,$Hz for maximum and minimum 
compactness $M/R$ for typical neutron star models. Shaded areas are excluded 
from the possible parameter space to explain the observations.}
\label{fig_mu_B}
\end{figure*}
We now combine our previous findings in order to constrain some properties 
of the magnetar. Plugging Eq.\,(\ref{eq:crustfreq_0}) into 
Eq.\,(\ref{eq:crustfreq}) we obtain a relation between $\mu_{cc}$ and $\bar B$: 
\begin{equation}
 c_s=\sqrt{\frac{\mu_{cc}}{\rho_{cc}}}\lesssim 
 c_s^\mathrm{ref} \frac{f_{^2t_n}}{f^\mathrm{ref}_{^2t_n^{(0)}}}
 \frac{1}{\sqrt{1+a_{2t_n} \bar B^2}}
 \frac{\Delta r}{\Delta r_\mathrm{ref}}
 \frac{\left(1-\frac{2M_\mathrm{ref}}{R_\mathrm{ref}}\right)}{ 
\left(1-\frac{2M }{R}\right)}\,.
\end{equation}
The 
corresponding plots 
for different crust sizes and for $M/R=0.14$ as the most constraining case are 
shown in the two panels of 
Fig.\,\ref{fig_mu_B} as blue ($n=1$) and green ($n=2$) lines and for 
$\varepsilon_\star X_c=0.046$
(left panel) and $\varepsilon_\star X_c=0.368$ (right panel). Higher shear 
moduli than those leading to the lines plotted in the figure are excluded, 
because the oscillation frequency would be too high.  The black line in 
each figure shows the limit of the minimum magnetic field above which the QPO 
breaks out of the surface, according to Eq.\,(\ref{eq:bbreak}). If the shear 
modulus and therefore $c_s$ are too strong, the oscillations stay confined in 
the core. The last constraint in Fig.\,\ref{fig_mu_B} is the matching of the 
low-frequency fundamental oscillation to $f_{^2U_2}=30\,$Hz 
(Eq.\,(\ref{eq:f2u2})) as indicated with red lines for $M/R=0.14$ and 
$M/R=0.28$. These values for $M/R$ are chosen because  they represent the 
maximum and minimum compactnesses of typical neutron star models as given in 
\cite{Sotani2008}. All shaded areas in Fig.\,\ref{fig_mu_B} indicate forbidden 
regions according to the previous constraints. In addition we also show in the 
figure the location in the $c_s-\bar B-$plane of our fiducial model with the DH 
EoS for the crust with the spin down estimate of the magnetic-field strength of 
$\bar B=2.1\times10^{15}\,$G for SGR 1806-20. 

From Fig.\,\ref{fig_mu_B} we conclude that for a realistic value of the 
entrainment factor, $\varepsilon_\star X_c=0.046$ (left panel), our model with 
the DH EoS for SGR 1806-20 is consistent with the observations only 
if the crust is quite extended, $\Delta r \gtrsim1.3\,$km. If we change the 
entrainment factor to scan the possible parameter space, we find consistent 
solutions only up to $\varepsilon_\star X_c=0.368$ (right panel). For higher 
values of $\varepsilon_\star X_c$ the star would be less compact $M/R<0.14$ than 
allowed by a realistic EoS. In other words, our model constrains the entrainment 
factor to be  $\varepsilon_\star X_c<0.368$.

%..........................................................................................................................
\subsection{Bayesian analysis}\label{sec_Bayes}

In order to test observational data $\D$ against our model $\M$, we perform in this section a Bayesian analysis. This analysis allows to describe the state of knowledge about an uncertain hypothesis model $\M$ and about the $N$ parameters $\theta_i$, $i=1, \dots, N$, of the model. Baye's theorem,
\begin{equation}
p ({\pmb\theta}|\D,\M) = \frac{p (\D|{\pmb \theta},\M) \, p ({\pmb \theta|\M})}{p(\D|\M)},
\end{equation}
allows to compute the {\it posterior} probability density function (PDF), $p ({\pmb\theta}|\D,\M)$, for a set of parameters $\pmb{\theta}=\{\theta_i \}$, given the model and the data. The {\it likelihood}, $p (\D|{\pmb \theta},\M)$, is the probability to observe certain data, given $\M$ and $\pmb \theta$, and can be computed straightforwardly from the model,
as we see next. The {\it prior} PDF of the parameters, $p ({\pmb \theta|\M})$, fulfills 
\begin{equation}
\int_\V d\theta_1\ldots d\theta_N \, p ({\pmb \theta|\M}) = 1\,, \label{eq:priornorm}
\end{equation}
i.e.~it is assumed that the parameters of the model only take values in a certain
range given by $\V$, a $N$-dimensional volume in the parameter space.
Finally, the {\it evidence}, $p(\D|\M)$, is the PDF to observe certain data
given a model, regardless of the parameters. It can be computed as
\begin{equation}
p(\D|\M) = \int_\V d\theta_1\ldots d\theta_N  \, p (\D|{\pmb \theta},\M) \, p ({\pmb \theta|\M}) \,.
\end{equation}

The resulting posterior PDF describes the collective knowledge about the different parameters and how they are related. Results for a specific subset of parameters, ${\pmb \theta}_{\rm sub} = \{ \theta_a,\theta_b...\}$, can be obtain by
marginalizing over the rest of the parameters, i.e.
\begin{equation}
p({\pmb \theta}_{\rm sub} |\D,\M) = \int \prod_{i=1, i \ne a,b,\ldots}^N d\theta_i \, p ({\pmb \theta}|\D,\M).
\end{equation}

In our model we consider that the observed QPOs are internal oscillations of the magnetar and make the next three model hypothesis:
\begin{itemize}
\item $\M_1$: The QPO frequency observed at about $30$~Hz, corresponds to the frequency
of the $^2U_2$ mode given by Eq.~(\ref{eq:f2u2}).

\item $\M_2$: The high-frequency QPO  observed at about $600$~Hz, corresponds to the frequency
of the magnetically modified $^2t_1$ QPO, and its frequency is 
larger or equal to that of the corresponding unmagnetized 
oscillation, given by Eq.~(\ref{eq:crustfreq}), and smaller than twice this value (Eq.~(\ref{eq:highmax})).

\item $\M_3$: QPOs are only observed if the magnetic field, $\bar{B} $, is 
higher than the breakout magnetic field given by Eq.~(\ref{eq:bbreak}) and 
lower than the maximum field given by Eq.~(\ref{eq:bmax}).

\end{itemize}
Moreover, this model consists of $N=5$ parameters, namely:
\begin{itemize}
\item $\bar B$, the magnetar magnetic field. We consider values of $\bar B$ in the range
$10^{13}$ to $10^{16}$~G, which is a reasonable range for magnetars from 
observational and theoretical expectations.
\item $\varepsilon_\star X_c$, which is a measure of the coupling between 
neutrons 
and protons in the core, and, hence, an indication for superfluidity. Possible 
values are in the range $0$ to $1$.
\item $c_s$, the shear speed at the base of the crust. We take values in the 
range $10^7$ to $2\times10^8$~cm/s, which contains all the values considered 
in \cite{Steiner2009} (ranging from $3\times10^7$ to $1.4\times10^8$~cm/s).
\item $\Delta r$, the crust thickness. We take values in the range  $0.3$ 
to $3$~km, which includes e.g. the range of values in \cite{Sotani2008} from 
$0.317$ to $1.654$ km.
\item $M/R$, the compactness of the star. Typical values for a variety of neutron star
masses and EoS give values in the range $0.1$ to $0.3$, which includes e.g. the range of 
values in \cite{Sotani2008} from $0.14$ to $0.28$.
\end{itemize}
Given a set of parameters ${\pmb \theta} = (\bar B, \varepsilon_\star X_c, c_s, 
\Delta r, M/R)$
it is possible to construct a model $\M = (\M_1, \M_2, \M_3)$ to compare with observations. 

To construct the prior, $p ({\pmb \theta|\M})$, we consider all possible values 
of $\varepsilon_\star X_c$, $c_s$ and $\Delta r$ equally probable within the 
parameter space.  For the dependence on $\bar B$, we construct the prior in 
order to fulfill model hypothesis $\M_3$:
\begin{eqnarray}
p ({\pmb \theta|\M}) &\propto& \frac{1}{2} \left(1 + \rm{erf} \left(\frac{\bar B - 
\bar B_{\rm breakout}}{ \sigma_{\rm sys, breakout} \sqrt{2}}\right)\right) 
\times \nonumber \\
&&  \frac{1}{2} \left(1 + \rm{erf} \left(\frac{\bar B_{\rm max} - \bar B}{\sigma_{\rm sys, max} \sqrt{2} }\right)\right) , 
\end{eqnarray}
where $\sigma_{\rm sys, breakout}$ and $\sigma_{\rm sys, max}$ correspond to 
the systematic error estimated in 
Section~\ref{sec_breakout} for $\bar B_{\rm breakout}$ and $\bar B_{\rm max}$, 
respectively. The normalization
constant is computed in order to fulfill  Eq.~(\ref{eq:priornorm}).

For our analysis we consider the RXTE observation of the 2004 giant flare in SGR 1806-20 \citep{Strohmayer2006}. 
We identify the $f_{^2U_2}$ frequency appearing in $\M_1$ as the QPO at $f^{\rm obs}_{^2U_2}=29.0\pm0.4$~Hz with a width 
$\sigma^{\rm obs}_{^2U_2}=4.1$~Hz
(data $\D_1$ hereafter), and the $f_{^2t_1}$ frequency appearing in $\M_2$ as the QPO at $f^{\rm obs}_{^2t_1}=625.5\pm0.2$~Hz with a width 
 $\sigma^{\rm obs}_{^2t_1}=1.8$~Hz 
(data $\D_2$ hereafter).  Using this observational data, $\D = \{ \D_1, \D_2\}$, it is possible to compute the likelihood as 
\begin{equation}
p (\D|{\pmb \theta},\M) = \prod_{i=1}^{N_O}p (\D_i|{\pmb \theta},\M),
\end{equation}
being $N_O=2$ the number of observational data points. We assume the data follow a Gaussian distribution,
\begin{eqnarray}
p (\D_1|{\pmb \theta},\M) &=& %\frac{1}{\sqrt{2\pi} \sigma_1 } 
\exp \left[  -\frac{1}{2} \left(\frac{f_{^2U_2} - f^{\rm obs}_{^2U_2}}{\sigma_1}\right)^2 \right],\\
p (\D_2|{\pmb \theta},\M) &=& \frac{1}{4} \left[ 1 - \rm{erf} \left(\frac{f^{\rm 
obs}_{^2t_1} - f_{^2t_1}}{\sqrt{2} \sigma_2}\right) \right] \cdot \nonumber \\
&& \left[ 1 - \rm{erf} \left(\frac{2 f_{^2t_1} - f^{\rm obs}_{^2t_1} }{\sqrt{2} \sigma_2}\right) \right],
\end{eqnarray}
where $\sigma_i$ is the combined uncertainty from the observation and the model prediction and can be estimated as 
\begin{equation}
\sigma_i = \sqrt{(\sigma^{\rm sys}_i)^2 + (\sigma^{\rm obs}_i)^2},
\end{equation}
where $\sigma^{\rm sys}_i$ and $\sigma^{\rm obs}_i$ correspond to systematic errors of the model and observations, respectively. For $\sigma^{\rm obs}_1$ and $\sigma^{\rm obs}_2$ we take the width of the observed QPOs described above.
Note that it is not necessary to normalize the likelihood $p (\D|{\pmb \theta},\M)$ to $1$, because any normalization constant cancels out in the computation of the posterior by the same constant appearing in the evidence, $p(\D|\M)$.

\begin{figure}
\includegraphics[width=.24\textwidth]{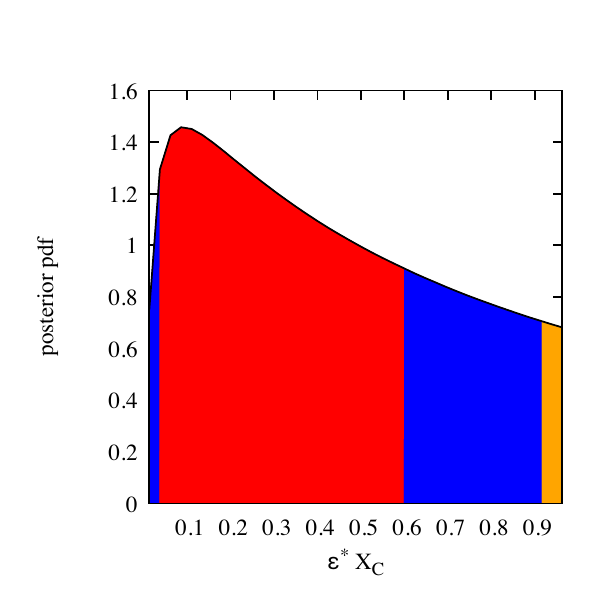}
\hspace{-0.2cm}
\includegraphics[width=.24\textwidth]{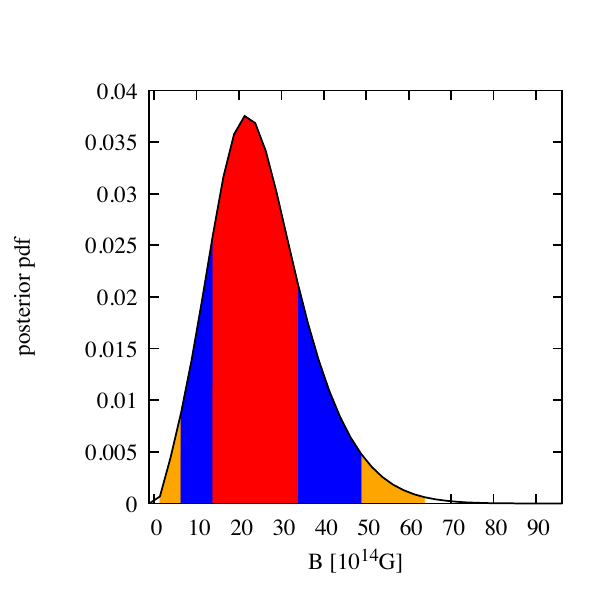}
\\
\includegraphics[width=.24\textwidth]{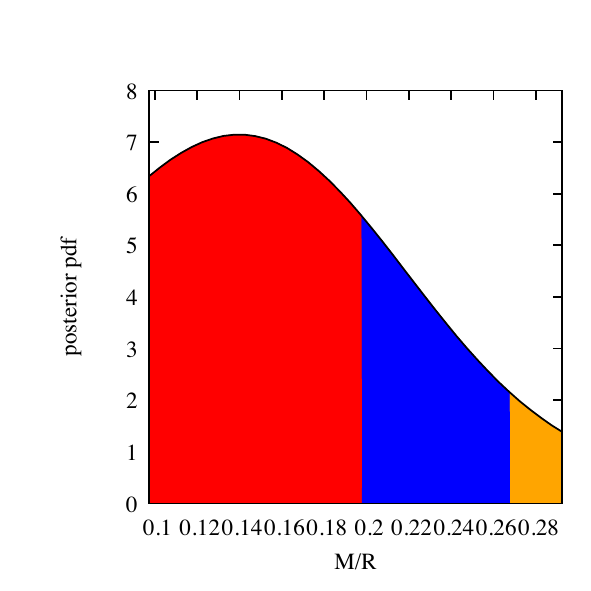}
\hspace{-0.2cm}
\includegraphics[width=.24\textwidth]{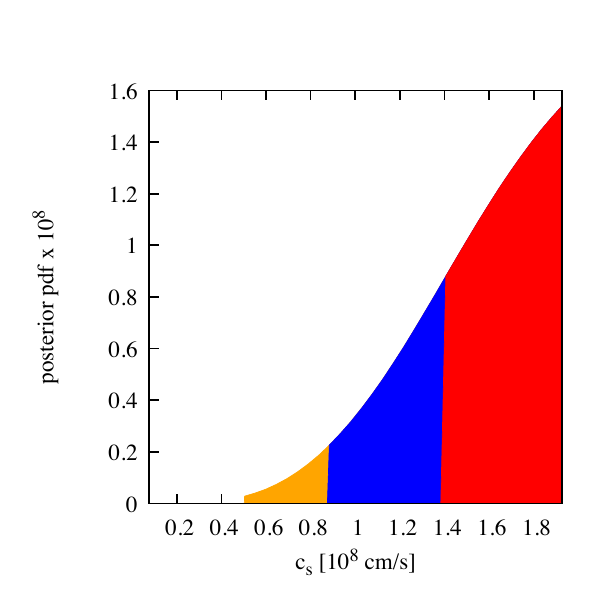}
\\
\includegraphics[width=.24\textwidth]{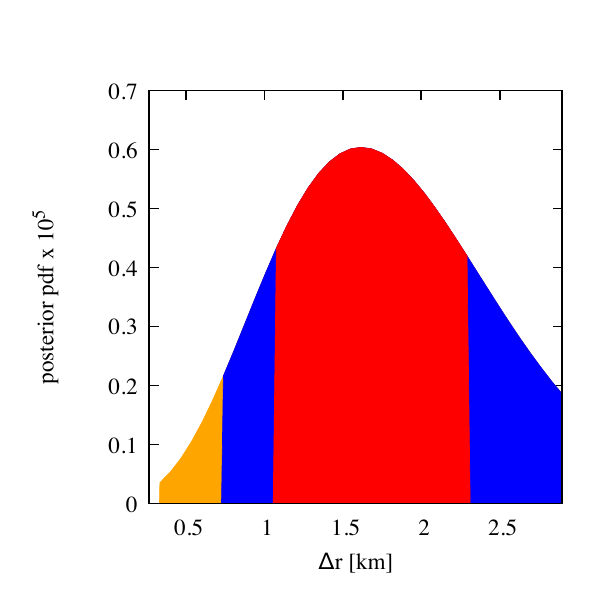}
\caption{1D marginalized posteriors for all  parameters (black line) and
credible intervals for $68.2\%$ (1 $\sigma$, red region), $95.4\%$ (2 $\sigma$, 
blue region)
and $99.7\%$ (3 $\sigma$, orange region).
}
\label{fig_post1d}
\end{figure}

Given that the model only depends on five parameters and the relative 
smallness of the parameter space, 
we can tackle the problem of the computation of the posteriors directly by
discretizing the 5-dimensional
parameter space in a grid. We use an equidistant grid for $\bar{B}$, 
$\varepsilon_\star X_c$, $c_s$, $\Delta r$ ans $M/R$ with $40$ points for each 
parameter. Once the posteriors are known we marginalize over unwanted parameters 
to obtain 1D plots of the posteriors for each parameter, and 2D plots for each 
possible couple of different parameters. Finally we obtain credible intervals by 
ordering the marginalized data in descending order of probability and 
integrating until the desired $\sigma$ level. 

\begin{table}
\setlength{\tabcolsep}{5pt}
\begin{tabular}{c c c c cc}
\hline
&$\varepsilon_\star X_c$ & $\bar B$ & $M/R$ & $c_s$ & 
$\Delta r$ \\ 
& & [$10^{14}$~G] & & [$10^8$~cm/s] &[km] \\ \hline
max. 
&$0.09^{+0.46}_{-0.07}$  
& $21^{+13}_{-10}$ 
 & $0.14^{+0.05}_{-0.04}$  
& $>1.4$
& $1.6^{+0.7}_{-0.6}$\\ 
median  
& 0.49
& 48.
& 0.20
& 1.0
& 1.6\\
\hline
\end{tabular}
\caption{Maximum values and 1-$\sigma$ intervals of the 1D marginalized 
posteriors and median values. For the case of $c_{\rm s}$, the maximum value 
is encountered at the boundary of the interval, so only a 1-$\sigma$ lower 
bound is given.}
\label{tab_median}
\end{table}

\begin{figure*}
\includegraphics[width=.95\textwidth]{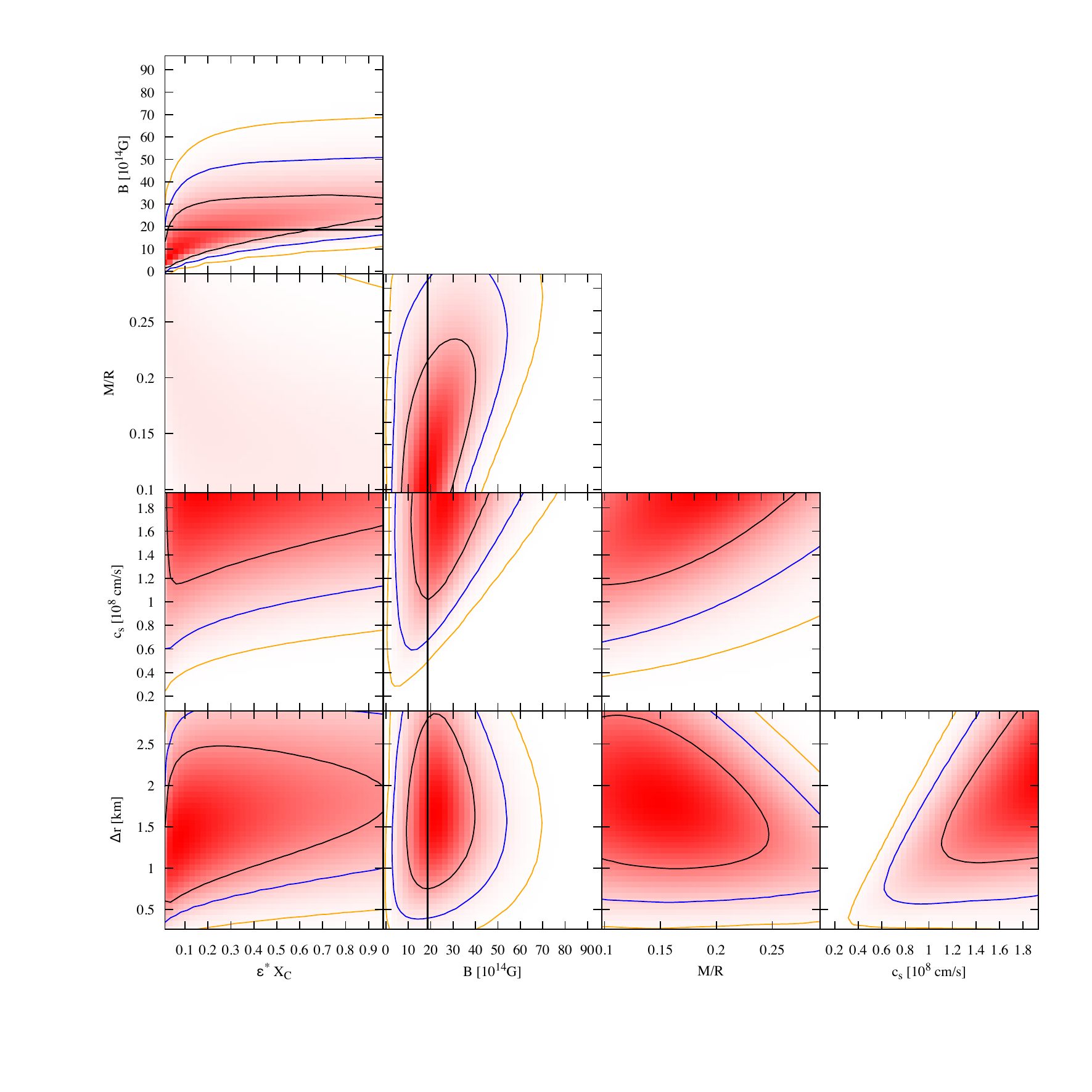}
\caption{2D marginalized posteriors for all possible combinations of parameters 
(color coded) and credible intervals for $68.2\%$ (1 $\sigma$, black line), 
$95.4\%$ (2 $\sigma$, blue line) and $99.7\%$ (3 $\sigma$, orange line). 
The thick black line indicates the spin-down estimate of the dipolar 
magnetic field strength of SGR 1806-20.
}
\label{fig_post2d}
\end{figure*}

Figs.~\ref{fig_post1d} and \ref{fig_post2d} show 1D and 2D marginalized 
posteriors for all parameters, respectively, while in Table~\ref{tab_median} we 
give the maximum and median values for each of the parameters based on the 
1D marginalized posteriors, including 1-$\sigma$ deviations. 
The posteriors are spread over a large region of the parameter space. To 
test what is causing the large errors in the posteriors, we performed an 
analysis reducing all systematic and 
observational errors to $1\%$ and obtained qualitatively similar results and 
errors in the posteriors. This implies that the errors are 
dominated
by the fact that we are using a model with $5$ parameters to fit only two 
observational data. Actually, in about $8\%$ of the volume of the parameter 
space we find QPO frequencies that match both observations with 1-$\sigma$ 
confidence. This indicates that there is a large degree of degeneracy in the 
parameter space, and hence
the posteriors produce only weak constraints.

The value of the magnetic-field strength for SGR 1806-20 based on our Bayesian 
analysis, $\bar B = 2.1^{+1.3}_{-1.0}\times10^{15}\,$G, is in remarkable 
agreement with its value based on the measured period spin-down $\bar B=2 \times 
10^{15}$~G \citep{Woods2007}. This may be an indication that the magnetic field 
is predominantly dipolar also inside the star and no 
high-order multipoles are present. The value of $\varepsilon_\star 
X_c=0.09^{+0.46}_{-0.07}$ is incompatible with a normal fluid core, i.e. we 
require a significant part of the matter to be superfluid in our model. In 
fact, the median value is close to the values theoretically predicted for completely 
superfluid models $\varepsilon_\star^0 X_c^0=0.046$ \citep{Douchin2001}.

Our analysis also favours a small compactness of $M/R\lesssim0.19$. Assuming 
a $1.4$~M$_\odot$ neutron star, this would imply a relatively stiff EoS. In 
addition, the mass of the object is likely to not be close to the maximum mass 
allowed by the EoS, because high-mass neutron stars are usually more compact. 
For example, if we
consider the additional constraints to the EOS by \cite{Lattimer2014}, our 
compactness restricts the mass of the neutron star to be in the interval 
$0.7-1.5~M_\odot$. Finally, both the inferred shear speed 
$c_s>1.4\times10^8\,$cm/s and crust thickness $\Delta 
r=1.61^{+0.7}_{-0.6}\,$km are on the upper range of the expectations of 
theoretical models \citep{Steiner2009,Sotani2008}. 

\subsection{Potential conflicts with observations}
The detection of very similar frequencies in the normal bursts of J1550-5418~\citep{Huppenkothen2014a} and in the giant flares of SGR 1806-20 and SGR 1900+14 poses a problem for their interpretation as magneto-elastic oscillations: Why do all three magnetars have very similar oscillations frequencies while their magnetic field strength estimates differ by almost one order of magnitude, $\bar B_{J1550}\sim 3.2\times10^{14}\,$G and $\bar B_{1806}\sim 2\times10^{15}\,$G? The frequencies of magneto-elastic oscillations are expected to scale with the magnetic-field strength like Eq.\,(\ref{eq:f2u2}). Our model thus 
predicts significantly higher frequencies for SGR 1806-20 than for SGR 1900+14, 
which in turn should have higher frequencies than J1550-5418. Indeed, SGR 
1900+14 ($28\,$Hz and $84\,$Hz) has slightly lower frequencies compared to SGR 
1806-20 ($30\,$Hz and $92\,$Hz). However, the difference is smaller than 
expected from Eq.\,(\ref{eq:f2u2_tmp}).
Within our model, the similar frequencies for different magnetic field strengths may still be explained by the following reasons:
i) According to Eq.\,(\ref{eq:f2u2}), different masses and thus different 
compactnesses of the neutron stars lead to different frequencies. The maximal 
allowed difference, depending on the compactness of the particular models, is a 
factor of roughly $2.5$. ii) Magnetic-field configurations different from the 
assumed dipolar-dominated configuration may influence the oscillations 
significantly. There may be strong toroidal fields whose influence on the 
oscillations have not been studied in sufficient detail yet. Higher order 
multipoles are unlikely, because they would shift the frequencies to too high 
values \citep{Gabler2013a}. iii) The spin-down estimate of the magnetic 
field strength is only accurate to a factor of 2 in the best case. iv) The 
similar frequencies may belong to different overtones and not to the 
fundamental oscillation. To break this degeneracy,  a clear identification of 
the observed magneto-elastic oscillation is necessary. Unfortunately, the 
observational data is yet insufficient to unambiguously determine which overtone 
or fundamental mode frequency can 
be associated with an observed frequency, i.e.~there is still no unambiguous pattern in the frequency spacing of the QPOs. Recent work on detecting QPOs in normal burst of magnetars is thus very promising \citep{Huppenkothen2014a,Huppenkothen2014b} and we strongly encourage further studies in this direction.

During the late phase of the writing of this paper a re-analysis of the 
observational data of the two giant flares of SGR 1806 and SGR 1900 was 
presented by \cite{Pumpe2017}. Unfortunately, they do not confirm the 
previously reported frequencies, but find new potential signals at $9.2\,$Hz 
and $7.7\,$Hz, respectively. They do not comment on the 
high-frequency QPOs above $150\,$Hz. Assuming that the newly found low 
frequency QPO at $9.2\,$Hz is the fundamental magneto-elastic QPO ($^2U_2$),
 we can perform perform the same Bayesian analysis to see how the results 
change with this different identification. We have considered two alternative 
cases, one in which we identify the $625$~Hz QPO as the high frequency QPO ($^2t_1$) 
and a second case in which we do not consider any high frequency QPO.
The results for the maximum values of the 1D marginalised posteriors are 
summarised and compared to our previous results in Table\,\ref{tab2}. 
The estimates for the magnetic field strength in 
all cases are still compatible with the spin-down estimate. With the lower 
frequency of $9.2\,$Hz, our estimates for the crust thickness and shear speed 
become lower than before, while the compactness becomes larger. 
$\varepsilon_\star X_c$ is significantly larger than previously 
estimated, but it is still consistent with a large fraction of the core being 
superfluid. Thus, it is of great importance to clarify, which frequencies are 
present in the signal.

\begin{table}
\setlength{\tabcolsep}{3pt}
\begin{tabular}{c c c c c cc}
\hline
$^2U_2$&$^2t_1$&$\varepsilon_\star X_c$ & $\bar B$ & $M/R$ & 
$c_s$ & $\Delta r$  \\
&&& [$10^{14}$~G] & & 
[$10^8$~cm/s] & [km]  \\ \hline
$29$~Hz &
$625$~Hz
&$0.09^{+0.46}_{-0.07}$  
& $21^{+13}_{-10}$ 
 & $0.14^{+0.05}_{-0.04}$  
& $>1.4$
& $1.6^{+0.7}_{-0.6}$\\ 
$9.2$~Hz &
$625$~Hz
&$0.5^{+0.4}_{-0.2}$  
& $14^{+10}_{-7}$ 
 & $>0.2$  
& $>1.2$
& $1.1^{+0.6}_{-0.5}$\\  
$9.2$~Hz &
-
&$0.4^{+0.4}_{-0.2}$  
& $11^{+10}_{-7}$ 
 & $>0.2$  
& $>1.0$
& -\\ \hline
\end{tabular}
\caption{Maximum values and 1-$\sigma$ intervals (or 1-$\sigma$ 
lower limits) of the 1D marginalised 
posteriors, for three different identifications of the QPOs in giant flares: the same as in table~\ref{tab_median} (upper row), 
using $9.2$~Hz as low frequency QPO instead of $29$~Hz (middle row), and  using only $9.2$~Hz (lower row). In the latter
case, there are no constrains on $\Delta r$.
}
\label{tab2}
\end{table}
\section{Conclusions}\label{sec_conclusion}

In this paper we have discussed torsional oscillations of highly magnetized 
neutron stars (magnetars). Those are described by means of two-dimensional, 
magneto-elastic-hydrodynamical simulations. Our model of superfluid 
magneto-elastic oscillations is able to explain both the low- and high-frequency 
QPOs observed in magnetars. In particular we have shown that the analysis of 
these oscillations provides constraints on the breakout magnetic-field 
strength, on their fundamental frequency, and on the frequency of a particularly 
excited overtone. We have also shown how to use this information to constraint  
properties of high-density matter in neutron stars in general, employing 
Bayesian analysis. 

Our model-dependent Bayesian posterior estimates for SGR 1806-20 give a 
magnetic-field strength $\bar B= 2.1^{+1.3}_{-1.0}\times10^{15}\,$G and 
a 
superfluid parameter $\varepsilon_\star X_c=0.09^{+0.46}_{-0.07}$, which 
encodes entrainment and proton fraction. The uncertainties given are 1-$\sigma$ 
error bars from the Bayesian inference. Both of these values agree remarkably 
well with observational and theoretical expectations, respectively. We have 
further 
obtained posterior estimates that favour a low value of the compactness of the 
star, $M/R\lesssim0.19$, indicating a relatively stiff EoS and/or low mass 
neutron star, a relatively high shear speed at the base of the crust,  
$c_s>1.4\times10^8\,$cm/s, and a thick crust with $\Delta 
r=1.61^{+0.7}_{-0.6}\,$km.

The main source of uncertainties in our analysis is neither the accuracy of our 
models
nor of the observations, but the lack of observational data. We are 
constraining five parameters
with only two observational data. As a consequence, there is a large degeneracy 
in the parameter
space, which produces large uncertainties in the posteriors. Future X-ray 
observatories could provide more data, either detecting QPOs in the more 
frequent intermediate flares or even looking to extra-galactic giant flares. It 
still has to be clarified how to stack data from different objects in a 
Bayesian analysis to produce meaningful and tight constraints on the properties 
of neutron star interiors. One of the problems is the use of parameters that 
depend on the particular object under consideration (magnetic field strength 
and configuration, and crust thickness $\Delta r$), which will
have to be substituted by general properties, which hold for all 
neutron stars with future QPO detections. This work is only a 
proof-of-principle that it is possible
to extract information from QPOs. The detailed multi-object analysis will be 
performed elsewhere.

Our model is however not complete, as it still does not include the possibility 
of the protons to be superconducting (which may be destroyed due to the 
ultra-strong magnetic fields in magnetars) nor the coupling of toroidal and 
poloidal oscillations, that may change the QPO spectrum. Therefore, there is 
room to improve our current interpretation of the QPOs observed in magnetar 
giant flares. Nevertheless, we have shown how a systematic analysis of the 
coupled magneto-elastic oscillations can help to constrain the properties of 
matter at nuclear densities. This is a significant improvement over previous 
studies that only took into account pure crustal shear modes. Furthermore, the 
constraints derived in this paper are general, in the sense that they  are based 
on general physical properties like the breakout of the oscillations, the 
matching of the fundamental oscillation, and the presence of a particular 
high-frequency QPO. Therefore, the procedure described here to obtain the 
constraints will still be valid in future improvements of the model (including 
e.g. superconductivity) and only the particular 
values of the quantities may change.

A main constraint derived in this work is the breakout of the oscillations from 
the core to the surface which depends on the magnetic-field strength and on the 
$\varepsilon_\star X_c$ parameter. By identifying the magnetic-field strength 
below which no QPOs are observed in magnetars, one could fix $\bar 
B_\mathrm{breakout}$ observationally and use Eq.\,(\ref{eq:bbreak}) to 
determine the superfluid properties ($\varepsilon_\star X_c$) for a given 
magnetic-field configuration. To make this kind of constraints possible, 
observations like those of \cite{Huppenkothen2014a,Huppenkothen2014b} are 
indispensable and more 
work in this direction is highly encouraged. Another important open question is how the oscillations of the neutron star may modulate the emission. To understand the modulation mechanism is essential to estimate the amplitude of the oscillations and to compare with the properties of the observed QPOs. We plan to address these issues in further work.

%===============================================================================
\section*{Acknowledgments}

Work supported by the Spanish MINECO (grant AYA2015-66899-C2-1-P),
 the {\it Generalitat Valenciana} (PROMETEOII-2014-069), and the EU through the 
ERC Starting Grant no. 259276-CAMAP and the ERC Advanced Grant
no. 341157-COCO2CASA. Partial support comes from the COST Actions 
 NewCompStar (MP1304) and PHAROS (CA16214). Computations were performed at the 
{\it Servei
  d'Inform\`atica de la Universitat de Val\`encia} and at the 
Max Planck Computing and Data Facility (MPCDF).

\bibliographystyle{mnras}
\bibliography{magnetar}

\begin{thebibliography}{}
\makeatletter
\relax
\def\mn@urlcharsother{\let\do\@makeother \do\$\do\&\do\#\do\^\do\_\do\%\do\~}
\def\mn@doi{\begingroup\mn@urlcharsother \@ifnextchar [ {\mn@doi@}
  {\mn@doi@[]}}
\def\mn@doi@[#1]#2{\def\@tempa{#1}\ifx\@tempa\@empty \href
  {http://dx.doi.org/#2} {doi:#2}\else \href {http://dx.doi.org/#2} {#1}\fi
  \endgroup}
\def\mn@eprint#1#2{\mn@eprint@#1:#2::\@nil}
\def\mn@eprint@arXiv#1{\href {http://arxiv.org/abs/#1} {{\tt arXiv:#1}}}
\def\mn@eprint@dblp#1{\href {http://dblp.uni-trier.de/rec/bibtex/#1.xml}
  {dblp:#1}}
\def\mn@eprint@#1:#2:#3:#4\@nil{\def\@tempa {#1}\def\@tempb {#2}\def\@tempc
  {#3}\ifx \@tempc \@empty \let \@tempc \@tempb \let \@tempb \@tempa \fi \ifx
  \@tempb \@empty \def\@tempb {arXiv}\fi \@ifundefined
  {mn@eprint@\@tempb}{\@tempb:\@tempc}{\expandafter \expandafter \csname
  mn@eprint@\@tempb\endcsname \expandafter{\@tempc}}}

\bibitem[\protect\citeauthoryear{{Akmal}, {Pandharipande}  \&
  {Ravenhall}}{{Akmal} et~al.}{1998}]{Akmal1998}
{Akmal} A.,  {Pandharipande} V.~R.,   {Ravenhall} D.~G.,  1998, \mn@doi [\prc]
  {10.1103/PhysRevC.58.1804}, \href
  {http://adsabs.harvard.edu/abs/1998PhRvC..58.1804A} {58, 1804}

\bibitem[\protect\citeauthoryear{{Andersson}, {Glampedakis}  \&
  {Samuelsson}}{{Andersson} et~al.}{2009}]{Andersson2009}
{Andersson} N.,  {Glampedakis} K.,   {Samuelsson} L.,  2009, \mn@doi [\mnras]
  {10.1111/j.1365-2966.2009.14734.x}, \href
  {http://adsabs.harvard.edu/abs/2009MNRAS.396..894A} {396, 894}

\bibitem[\protect\citeauthoryear{{Bocquet}, {Bonazzola}, {Gourgoulhon}  \&
  {Novak}}{{Bocquet} et~al.}{1995}]{Bocquet1995}
{Bocquet} M.,  {Bonazzola} S.,  {Gourgoulhon} E.,   {Novak} J.,  1995, \aap,
  \href {http://adsabs.harvard.edu/abs/1995A%26A...301..757B} {301, 757}

\bibitem[\protect\citeauthoryear{{Cerd{\'a}-Dur{\'a}n}, {Font}, {Ant{\'o}n}  \&
  {M{\"u}ller}}{{Cerd{\'a}-Dur{\'a}n} et~al.}{2008}]{Cerda2008}
{Cerd{\'a}-Dur{\'a}n} P.,  {Font} J.~A.,  {Ant{\'o}n} L.,   {M{\"u}ller} E.,
  2008, \mn@doi [\aap] {10.1051/0004-6361:200810086}, \href
  {http://adsabs.harvard.edu/abs/2008A%26A...492..937C} {492, 937}

\bibitem[\protect\citeauthoryear{{Cerd{\'a}-Dur{\'a}n}, {Stergioulas}  \&
  {Font}}{{Cerd{\'a}-Dur{\'a}n} et~al.}{2009}]{Cerda2009}
{Cerd{\'a}-Dur{\'a}n} P.,  {Stergioulas} N.,   {Font} J.~A.,  2009, \mn@doi
  [\mnras] {10.1111/j.1365-2966.2009.15056.x}, \href
  {http://adsabs.harvard.edu/abs/2009MNRAS.397.1607C} {397, 1607}

\bibitem[\protect\citeauthoryear{{Chamel}}{{Chamel}}{2012}]{Chamel2012}
{Chamel} N.,  2012, \mn@doi [\prc] {10.1103/PhysRevC.85.035801}, \href
  {http://adsabs.harvard.edu/abs/2012PhRvC..85c5801C} {85, 035801}

\bibitem[\protect\citeauthoryear{Chamel \& Haensel}{Chamel \&
  Haensel}{2008}]{Chamel2008}
Chamel N.,  Haensel P.,  2008, Living Reviews in Relativity, 11

\bibitem[\protect\citeauthoryear{{Colaiuda} \& {Kokkotas}}{{Colaiuda} \&
  {Kokkotas}}{2011}]{Colaiuda2011}
{Colaiuda} A.,  {Kokkotas} K.~D.,  2011, \mn@doi [\mnras]
  {10.1111/j.1365-2966.2011.18602.x}, \href
  {http://adsabs.harvard.edu/abs/2011MNRAS.414.3014C} {414, 3014}

\bibitem[\protect\citeauthoryear{{Colaiuda}, {Beyer}  \& {Kokkotas}}{{Colaiuda}
  et~al.}{2009}]{Colaiuda2009}
{Colaiuda} A.,  {Beyer} H.,   {Kokkotas} K.~D.,  2009, \mn@doi [\mnras]
  {10.1111/j.1365-2966.2009.14878.x}, \href
  {http://adsabs.harvard.edu/abs/2009MNRAS.396.1441C} {396, 1441}

\bibitem[\protect\citeauthoryear{{Deibel}, {Steiner}  \& {Brown}}{{Deibel}
  et~al.}{2014}]{Deibel2014}
{Deibel} A.~T.,  {Steiner} A.~W.,   {Brown} E.~F.,  2014, \mn@doi [\prc]
  {10.1103/PhysRevC.90.025802}, \href
  {http://adsabs.harvard.edu/abs/2014PhRvC..90b5802D} {90, 025802}

\bibitem[\protect\citeauthoryear{{Douchin} \& {Haensel}}{{Douchin} \&
  {Haensel}}{2001}]{Douchin2001}
{Douchin} F.,  {Haensel} P.,  2001, \mn@doi [\aap]
  {10.1051/0004-6361:20011402}, \href
  {http://adsabs.harvard.edu/abs/2001A%26A...380..151D} {380, 151}

\bibitem[\protect\citeauthoryear{{Duncan}}{{Duncan}}{1998}]{Duncan1998}
{Duncan} R.~C.,  1998, \mn@doi [\apjl] {10.1086/311303}, \href
  {http://adsabs.harvard.edu/abs/1998ApJ...498L..45D} {498, L45}

\bibitem[\protect\citeauthoryear{{Gabler}, {Cerd{\'a} Dur{\'a}n}, {Font},
  {M{\"u}ller}  \& {Stergioulas}}{{Gabler} et~al.}{2011}]{Gabler2011letter}
{Gabler} M.,  {Cerd{\'a} Dur{\'a}n} P.,  {Font} J.~A.,  {M{\"u}ller} E.,
  {Stergioulas} N.,  2011, \mn@doi [\mnras] {10.1111/j.1745-3933.2010.00974.x},
  \href {http://adsabs.harvard.edu/abs/2011MNRAS.410L..37G} {410, L37}

\bibitem[\protect\citeauthoryear{{Gabler}, {Cerd{\'a}-Dur{\'a}n},
  {Stergioulas}, {Font}  \& {M{\"u}ller}}{{Gabler} et~al.}{2012}]{Gabler2012}
{Gabler} M.,  {Cerd{\'a}-Dur{\'a}n} P.,  {Stergioulas} N.,  {Font} J.~A.,
  {M{\"u}ller} E.,  2012, \mn@doi [\mnras] {10.1111/j.1365-2966.2012.20454.x},
  \href {http://adsabs.harvard.edu/abs/2012MNRAS.421.2054G} {421, 2054}

\bibitem[\protect\citeauthoryear{{Gabler}, {Cerd{\'a}-Dur{\'a}n},
  {Stergioulas}, {Font}  \& {M{\"u}ller}}{{Gabler} et~al.}{2013a}]{Gabler2013b}
{Gabler} M.,  {Cerd{\'a}-Dur{\'a}n} P.,  {Stergioulas} N.,  {Font} J.~A.,
  {M{\"u}ller} E.,  2013a, \mn@doi [Physical Review Letters]
  {10.1103/PhysRevLett.111.211102}, \href
  {http://adsabs.harvard.edu/abs/2013PhRvL.111u1102G} {111, 211102}

\bibitem[\protect\citeauthoryear{{Gabler}, {Cerd{\'a}-Dur{\'a}n}, {Font},
  {M{\"u}ller}  \& {Stergioulas}}{{Gabler} et~al.}{2013b}]{Gabler2013a}
{Gabler} M.,  {Cerd{\'a}-Dur{\'a}n} P.,  {Font} J.~A.,  {M{\"u}ller} E.,
  {Stergioulas} N.,  2013b, \mn@doi [\mnras] {10.1093/mnras/sts721}, \href
  {http://adsabs.harvard.edu/abs/2013MNRAS.430.1811G} {430, 1811}

\bibitem[\protect\citeauthoryear{{Gabler}, {Cerd{\'a}-Dur{\'a}n}, {Font},
  {Stergioulas}  \& {M{\"u}ller}}{{Gabler} et~al.}{2014a}]{Gabler2014a}
{Gabler} M.,  {Cerd{\'a}-Dur{\'a}n} P.,  {Font} J.~A.,  {Stergioulas} N.,
  {M{\"u}ller} E.,  2014a, \mn@doi [Astronomische Nachrichten]
  {10.1002/asna.201312025}, \href
  {http://adsabs.harvard.edu/abs/2014AN....335..240G} {335, 240}

\bibitem[\protect\citeauthoryear{{Gabler}, {Cerd{\'a}-Dur{\'a}n},
  {Stergioulas}, {Font}  \& {M{\"u}ller}}{{Gabler} et~al.}{2014b}]{Gabler2014b}
{Gabler} M.,  {Cerd{\'a}-Dur{\'a}n} P.,  {Stergioulas} N.,  {Font} J.~A.,
  {M{\"u}ller} E.,  2014b, \mn@doi [\mnras] {10.1093/mnras/stu1263}, \href
  {http://adsabs.harvard.edu/abs/2014MNRAS.443.1416G} {443, 1416}

\bibitem[\protect\citeauthoryear{{Gabler}, {Cerd{\'a}-Dur{\'a}n},
  {Stergioulas}, {Font}  \& {M{\"u}ller}}{{Gabler} et~al.}{2016}]{Gabler2016}
{Gabler} M.,  {Cerd{\'a}-Dur{\'a}n} P.,  {Stergioulas} N.,  {Font} J.~A.,
  {M{\"u}ller} E.,  2016, \mn@doi [\mnras] {10.1093/mnras/stw1272}, \href
  {http://adsabs.harvard.edu/abs/2016MNRAS.460.4242G} {460, 4242}

\bibitem[\protect\citeauthoryear{{Glampedakis}, {Samuelsson}  \&
  {Andersson}}{{Glampedakis} et~al.}{2006}]{Glampedakis2006b}
{Glampedakis} K.,  {Samuelsson} L.,   {Andersson} N.,  2006, \mn@doi [\mnras]
  {10.1111/j.1745-3933.2006.00211.x}, \href
  {http://adsabs.harvard.edu/abs/2006MNRAS.371L..74G} {371, L74}

\bibitem[\protect\citeauthoryear{{Glampedakis}, {Andersson}  \&
  {Samuelsson}}{{Glampedakis} et~al.}{2011}]{Glampedakis2011a}
{Glampedakis} K.,  {Andersson} N.,   {Samuelsson} L.,  2011, \mn@doi [\mnras]
  {10.1111/j.1365-2966.2010.17484.x}, \href
  {http://adsabs.harvard.edu/abs/2011MNRAS.410..805G} {410, 805}

\bibitem[\protect\citeauthoryear{{Hambaryan}, {Neuh{\"a}user}  \&
  {Kokkotas}}{{Hambaryan} et~al.}{2011}]{Hambaryan2011}
{Hambaryan} V.,  {Neuh{\"a}user} R.,   {Kokkotas} K.~D.,  2011, \mn@doi [\aap]
  {10.1051/0004-6361/201015273}, \href
  {http://adsabs.harvard.edu/abs/2011A%26A...528A..45H} {528, A45+}

\bibitem[\protect\citeauthoryear{{Huppenkothen} et~al.,}{{Huppenkothen}
  et~al.}{2014a}]{Huppenkothen2014a}
{Huppenkothen} D.,  et~al., 2014a, \mn@doi [\apj]
  {10.1088/0004-637X/787/2/128}, \href
  {http://adsabs.harvard.edu/abs/2014ApJ...787..128H} {787, 128}

\bibitem[\protect\citeauthoryear{{Huppenkothen}, {Watts}  \&
  {Levin}}{{Huppenkothen} et~al.}{2014b}]{Huppenkothen2014c}
{Huppenkothen} D.,  {Watts} A.~L.,   {Levin} Y.,  2014b, \mn@doi [\apj]
  {10.1088/0004-637X/793/2/129}, \href
  {http://adsabs.harvard.edu/abs/2014ApJ...793..129H} {793, 129}

\bibitem[\protect\citeauthoryear{{Huppenkothen}, {Heil}, {Watts}  \& {G{\"o}{\u
  g}{\"u}{\c s}}}{{Huppenkothen} et~al.}{2014c}]{Huppenkothen2014b}
{Huppenkothen} D.,  {Heil} L.~M.,  {Watts} A.~L.,   {G{\"o}{\u g}{\"u}{\c s}}
  E.,  2014c, \mn@doi [\apj] {10.1088/0004-637X/795/2/114}, \href
  {http://adsabs.harvard.edu/abs/2014ApJ...795..114H} {795, 114}

\bibitem[\protect\citeauthoryear{{Israel} et~al.,}{{Israel}
  et~al.}{2005}]{Israel2005}
{Israel} G.~L.,  et~al., 2005, \mn@doi [\apjl] {10.1086/432615}, \href
  {http://adsabs.harvard.edu/abs/2005ApJ...628L..53I} {628, L53}

\bibitem[\protect\citeauthoryear{{Lattimer} \& {Steiner}}{{Lattimer} \&
  {Steiner}}{2014}]{Lattimer2014}
{Lattimer} J.~M.,  {Steiner} A.~W.,  2014, \mn@doi [\apj]
  {10.1088/0004-637X/784/2/123}, \href
  {http://adsabs.harvard.edu/abs/2014ApJ...784..123L} {784, 123}

\bibitem[\protect\citeauthoryear{{Levin}}{{Levin}}{2006}]{Levin2006}
{Levin} Y.,  2006, \mn@doi [\mnras] {10.1111/j.1745-3933.2006.00155.x}, \href
  {http://adsabs.harvard.edu/abs/2006MNRAS.368L..35L} {368, L35}

\bibitem[\protect\citeauthoryear{{Levin}}{{Levin}}{2007}]{Levin2007}
{Levin} Y.,  2007, \mn@doi [\mnras] {10.1111/j.1365-2966.2007.11582.x}, \href
  {http://adsabs.harvard.edu/abs/2007MNRAS.377..159L} {377, 159}

\bibitem[\protect\citeauthoryear{{Messios}, {Papadopoulos}  \&
  {Stergioulas}}{{Messios} et~al.}{2001}]{Messios2001}
{Messios} N.,  {Papadopoulos} D.~B.,   {Stergioulas} N.,  2001, \mn@doi
  [\mnras] {10.1046/j.1365-8711.2001.04645.x}, \href
  {http://adsabs.harvard.edu/abs/2001MNRAS.328.1161M} {328, 1161}

\bibitem[\protect\citeauthoryear{{Passamonti} \& {Lander}}{{Passamonti} \&
  {Lander}}{2013}]{Passamonti2013}
{Passamonti} A.,  {Lander} S.~K.,  2013, \mn@doi [\mnras]
  {10.1093/mnras/sts372}, \href
  {http://adsabs.harvard.edu/abs/2013MNRAS.429..767P} {429, 767}

\bibitem[\protect\citeauthoryear{{Passamonti} \& {Lander}}{{Passamonti} \&
  {Lander}}{2014}]{Passamonti2014}
{Passamonti} A.,  {Lander} S.~K.,  2014, \mn@doi [\mnras]
  {10.1093/mnras/stt2134}, \href
  {http://adsabs.harvard.edu/abs/2014MNRAS.438..156P} {438, 156}

\bibitem[\protect\citeauthoryear{{Passamonti} \& {Pons}}{{Passamonti} \&
  {Pons}}{2016}]{Passamonti2016}
{Passamonti} A.,  {Pons} J.~A.,  2016, \mn@doi [\mnras]
  {10.1093/mnras/stw1880}, \href
  {http://adsabs.harvard.edu/abs/2016MNRAS.463.1173P} {463, 1173}

\bibitem[\protect\citeauthoryear{{Piro}}{{Piro}}{2005}]{Piro2005}
{Piro} A.~L.,  2005, \mn@doi [\apjl] {10.1086/499049}, \href
  {http://adsabs.harvard.edu/abs/2005ApJ...634L.153P} {634, L153}

\bibitem[\protect\citeauthoryear{{Pumpe}, {Gabler}, {Steininger}  \&
  {En{\ss}lin}}{{Pumpe} et~al.}{2017}]{Pumpe2017}
{Pumpe} D.,  {Gabler} M.,  {Steininger} T.,   {En{\ss}lin} T.~A.,  2017,
  preprint, \href {http://adsabs.harvard.edu/abs/2017arXiv170805702P} {}
  (\mn@eprint {arXiv} {1708.05702})

\bibitem[\protect\citeauthoryear{{Samuelsson} \& {Andersson}}{{Samuelsson} \&
  {Andersson}}{2007}]{Samuelsson2007}
{Samuelsson} L.,  {Andersson} N.,  2007, \mn@doi [\mnras]
  {10.1111/j.1365-2966.2006.11147.x}, \href
  {http://adsabs.harvard.edu/abs/2007MNRAS.374..256S} {374, 256}

\bibitem[\protect\citeauthoryear{{Samuelsson} \& {Andersson}}{{Samuelsson} \&
  {Andersson}}{2009}]{Samuelsson2009}
{Samuelsson} L.,  {Andersson} N.,  2009, \mn@doi [Classical and Quantum
  Gravity] {10.1088/0264-9381/26/15/155016}, \href
  {http://adsabs.harvard.edu/abs/2009CQGra..26o5016S} {26, 155016}

\bibitem[\protect\citeauthoryear{{Sedrakian}}{{Sedrakian}}{2016}]{Sedrakian2016}
{Sedrakian} A.,  2016, preprint, \href
  {http://adsabs.harvard.edu/abs/2016arXiv160100056S} {} (\mn@eprint {arXiv}
  {1601.00056})

\bibitem[\protect\citeauthoryear{{Sotani}}{{Sotani}}{2011}]{Sotani2011}
{Sotani} H.,  2011, \mn@doi [\mnras] {10.1111/j.1745-3933.2011.01122.x}, \href
  {http://adsabs.harvard.edu/abs/2011MNRAS.417L..70S} {417, L70}

\bibitem[\protect\citeauthoryear{{Sotani}, {Kokkotas}  \&
  {Stergioulas}}{{Sotani} et~al.}{2007}]{Sotani2007}
{Sotani} H.,  {Kokkotas} K.~D.,   {Stergioulas} N.,  2007, \mn@doi [\mnras]
  {10.1111/j.1365-2966.2006.11304.x}, \href
  {http://adsabs.harvard.edu/abs/2007MNRAS.375..261S} {375, 261}

\bibitem[\protect\citeauthoryear{{Sotani}, {Kokkotas}  \&
  {Stergioulas}}{{Sotani} et~al.}{2008}]{Sotani2008}
{Sotani} H.,  {Kokkotas} K.~D.,   {Stergioulas} N.,  2008, \mn@doi [\mnras]
  {10.1111/j.1745-3933.2007.00420.x}, \href
  {http://adsabs.harvard.edu/abs/2008MNRAS.385L...5S} {385, L5}

\bibitem[\protect\citeauthoryear{{Sotani}, {Nakazato}, {Iida}  \&
  {Oyamatsu}}{{Sotani} et~al.}{2012}]{Sotani2012}
{Sotani} H.,  {Nakazato} K.,  {Iida} K.,   {Oyamatsu} K.,  2012, \mn@doi
  [Physical Review Letters] {10.1103/PhysRevLett.108.201101}, \href
  {http://adsabs.harvard.edu/abs/2012PhRvL.108t1101S} {108, 201101}

\bibitem[\protect\citeauthoryear{{Sotani}, {Nakazato}, {Iida}  \&
  {Oyamatsu}}{{Sotani} et~al.}{2013a}]{Sotani2013}
{Sotani} H.,  {Nakazato} K.,  {Iida} K.,   {Oyamatsu} K.,  2013a, \mn@doi
  [\mnras] {10.1093/mnrasl/sls006}, \href
  {http://adsabs.harvard.edu/abs/2013MNRAS.428L..21S} {428, L21}

\bibitem[\protect\citeauthoryear{{Sotani}, {Nakazato}, {Iida}  \&
  {Oyamatsu}}{{Sotani} et~al.}{2013b}]{Sotani2013b}
{Sotani} H.,  {Nakazato} K.,  {Iida} K.,   {Oyamatsu} K.,  2013b, \mn@doi
  [\mnras] {10.1093/mnras/stt1152}, \href
  {http://adsabs.harvard.edu/abs/2013MNRAS.434.2060S} {434, 2060}

\bibitem[\protect\citeauthoryear{{Sotani}, {Iida}  \& {Oyamatsu}}{{Sotani}
  et~al.}{2016}]{Sotani2016}
{Sotani} H.,  {Iida} K.,   {Oyamatsu} K.,  2016, \mn@doi [\na]
  {10.1016/j.newast.2015.08.003}, \href
  {http://adsabs.harvard.edu/abs/2016NewA...43...80S} {43, 80}

\bibitem[\protect\citeauthoryear{{Steiner} \& {Watts}}{{Steiner} \&
  {Watts}}{2009}]{Steiner2009}
{Steiner} A.~W.,  {Watts} A.~L.,  2009, \mn@doi [Physical Review Letters]
  {10.1103/PhysRevLett.103.181101}, \href
  {http://adsabs.harvard.edu/abs/2009PhRvL.103r1101S} {103, 181101}

\bibitem[\protect\citeauthoryear{{Stergioulas} \& {Friedman}}{{Stergioulas} \&
  {Friedman}}{1995}]{Stergioulas1995}
{Stergioulas} N.,  {Friedman} J.~L.,  1995, \mn@doi [\apj] {10.1086/175605},
  \href {http://adsabs.harvard.edu/abs/1995ApJ...444..306S} {444, 306}

\bibitem[\protect\citeauthoryear{{Strohmayer} \& {Watts}}{{Strohmayer} \&
  {Watts}}{2005}]{Strohmayer2005}
{Strohmayer} T.~E.,  {Watts} A.~L.,  2005, \mn@doi [\apjl] {10.1086/497911},
  \href {http://adsabs.harvard.edu/abs/2005ApJ...632L.111S} {632, L111}

\bibitem[\protect\citeauthoryear{{Strohmayer} \& {Watts}}{{Strohmayer} \&
  {Watts}}{2006}]{Strohmayer2006}
{Strohmayer} T.~E.,  {Watts} A.~L.,  2006, \mn@doi [\apj] {10.1086/508703},
  \href {http://adsabs.harvard.edu/abs/2006ApJ...653..593S} {653, 593}

\bibitem[\protect\citeauthoryear{{Watts} \& {Strohmayer}}{{Watts} \&
  {Strohmayer}}{2006}]{Watts2006}
{Watts} A.~L.,  {Strohmayer} T.~E.,  2006, \mn@doi [\apjl] {10.1086/500735},
  \href {http://adsabs.harvard.edu/abs/2006ApJ...637L.117W} {637, L117}

\bibitem[\protect\citeauthoryear{{Woods}, {Kouveliotou}, {Finger}, {G{\"o}{\v
  g}{\"u}{\c s}}, {Wilson}, {Patel}, {Hurley}  \& {Swank}}{{Woods}
  et~al.}{2007}]{Woods2007}
{Woods} P.~M.,  {Kouveliotou} C.,  {Finger} M.~H.,  {G{\"o}{\v g}{\"u}{\c s}}
  E.,  {Wilson} C.~A.,  {Patel} S.~K.,  {Hurley} K.,   {Swank} J.~H.,  2007,
  \mn@doi [\apj] {10.1086/507459}, \href
  {http://adsabs.harvard.edu/abs/2007ApJ...654..470W} {654, 470}

\bibitem[\protect\citeauthoryear{{van Hoven} \& {Levin}}{{van Hoven} \&
  {Levin}}{2011}]{vanHoven2011}
{van Hoven} M.,  {Levin} Y.,  2011, \mn@doi [\mnras]
  {10.1111/j.1365-2966.2010.17499.x}, \href
  {http://adsabs.harvard.edu/abs/2011MNRAS.410.1036V} {410, 1036}

\bibitem[\protect\citeauthoryear{{van Hoven} \& {Levin}}{{van Hoven} \&
  {Levin}}{2012}]{vanHoven2012}
{van Hoven} M.,  {Levin} Y.,  2012, \mn@doi [\mnras]
  {10.1111/j.1365-2966.2011.20177.x}, \href
  {http://adsabs.harvard.edu/abs/2012MNRAS.420.3035V} {420, 3035}

\makeatother
\end{thebibliography}

\end{document}